\documentclass[twocolumn,times,tighten]{aastex63}

\usepackage{xspace}     
\usepackage{multirow}
\newcommand{\chandra}{\emph{Chandra}\xspace}
\newcommand{\xmmn}{\emph{XMM-Newton}\xspace}
\newcommand{\xmm}{\emph{XMM}\xspace}
\newcommand{\nustar}{\emph{NuSTAR}\xspace}
\newcommand{\galex}{\emph{GALEX}\xspace}
\newcommand{\wise}{\emph{WISE}\xspace}
\newcommand{\xdqso}{XDQSO\emph{z}\xspace}
\newcommand{\chanmaster}{\textsc{chanmaster}\xspace}

\newcommand{\borus}{\textsc{borus}\xspace}
\newcommand{\chisq}{$\chi^{\text{2}}$\xspace}
\newcommand{\chisqred}{$\chi_{\text{red}}^{\text{2}}$\xspace}
\newcommand{\ebv}{$E(\bv)_{\text{AGN}}$\xspace}
\newcommand{\w}[1]{\emph{W}#1\xspace}
\newcommand{\nh}{$N_{\text{H}}$\xspace}
\newcommand{\lx}{$L_{\text{X}}\xspace$}
\newcommand{\lxlim}{$L_{\text{X,lim}}$\xspace}
\newcommand{\lir}{$L_{\text{MIR}}$\xspace}
\newcommand{\lxir}{$L_{\text{X}}(L_{\text{MIR}})$\xspace}
\newcommand{\rlum}{$R_{L_{\text{X}}}$\xspace}
\newcommand{\llim}{$L_{{\text{X}}_{\text{lim}}}$\xspace}

\newcommand{\mum}{\,\mu{\text{m}}\xspace}
\newcommand{\cmcm}{cm$^{-{\text{2}}}$\xspace}
\newcommand{\ninit}{3,108,935\xspace}
\newcommand{\nisdss}{3,062,042\xspace}

\newcommand{\nixdqso}{46,893\xspace}

\newcommand{\niwise}{3,108,935\xspace}
\newcommand{\niunwise}{3,108,723\xspace}

\newcommand{\nitwom}{176,692\xspace}

\newcommand{\nigalex}{322,527\xspace}

\newcommand{\nicha}{365,678\xspace}
\newcommand{\nixmm}{2,888,356\xspace}
\newcommand{\ninst}{45,985\xspace}

\newcommand{\nzgood}{1,600,931\xspace}
\newcommand{\nzgoodtotf}{48.5\%\xspace}
\newcommand{\nrzgood}{1,600,931\xspace}
\newcommand{\nzgoodcumf}{48.5\%\xspace}
\newcommand{\nclean}{2,398,034\xspace}
\newcommand{\ncleantotf}{22.9\%\xspace}
\newcommand{\nrclean}{1,458,103\xspace}
\newcommand{\ncleancumf}{53.1\%\xspace}
\newcommand{\nbands}{1,407,768\xspace}
\newcommand{\nbandstotf}{54.7\%\xspace}
\newcommand{\nrbands}{960,373\xspace}
\newcommand{\nbandscumf}{69.1\%\xspace}
\newcommand{\nfourw}{98,186\xspace}
\newcommand{\nfourwtotf}{96.8\%\xspace}
\newcommand{\nrfourw}{60,994\xspace}
\newcommand{\nfourwcumf}{98.0\%\xspace}
\newcommand{\nnomsk}{2,108,427\xspace}
\newcommand{\nnomsktotf}{32.2\%\xspace}
\newcommand{\nrnomsk}{40,944\xspace}
\newcommand{\nnomskcumf}{98.7\%\xspace}
\newcommand{\nnodup}{3,108,878\xspace}
\newcommand{\nnoduptotf}{\textless0.01\%\xspace}
\newcommand{\nrnodup}{40,935\xspace}
\newcommand{\nnodupcumf}{98.7\%\xspace}
\newcommand{\nfinal}{40,935\xspace}
\newcommand{\nsdss}{40,349\xspace}

\newcommand{\nxdqso}{586\xspace}

\newcommand{\nwise}{40,935\xspace}
\newcommand{\nunwise}{40,934\xspace}

\newcommand{\ntwom}{15,236\xspace}

\newcommand{\ngalex}{18,681\xspace}
\newcommand{\ngal}{7,480\xspace}
\newcommand{\nagn}{33,455\xspace}
\newcommand{\ndet}{4,083\xspace}
\newcommand{\nnon}{36,852\xspace}
\newcommand{\nqgal}{0\xspace}
\newcommand{\nqagn}{3,483\xspace}
\newcommand{\nqdet}{317\xspace}
\newcommand{\nqnon}{3,166\xspace}
\newcommand{\ninfwac}{4,729\xspace}
\newcommand{\ndetwac}{1,820\xspace}
\newcommand{\nnonwac}{2,909\xspace}
\newcommand{\nqinfwac}{541\xspace}
\newcommand{\nqdetwac}{216\xspace}
\newcommand{\nqnonwac}{325\xspace}
\newcommand{\ninfcha}{5,939\xspace}
\newcommand{\ndetcha}{595\xspace}
\newcommand{\nnoncha}{5,344\xspace}
\newcommand{\nqinfcha}{1,197\xspace}
\newcommand{\nqdetcha}{97\xspace}
\newcommand{\nqnoncha}{1,100\xspace}
\newcommand{\ninfxmm}{37,146\xspace}
\newcommand{\ndetxmm}{3,424\xspace}
\newcommand{\nnonxmm}{33,722\xspace}
\newcommand{\nqinfxmm}{2,239\xspace}
\newcommand{\nqdetxmm}{216\xspace}
\newcommand{\nqnonxmm}{2,023\xspace}
\newcommand{\ninfnst}{701\xspace}
\newcommand{\ndetnst}{38\xspace}
\newcommand{\nnonnst}{663\xspace}
\newcommand{\nqinfnst}{47\xspace}
\newcommand{\nqdetnst}{4\xspace}
\newcommand{\nqnonnst}{43\xspace}
\newcommand{\nsrc}{40,935\xspace}
\newcommand{\nqwac}{541\xspace}
\newcommand{\nqres}{2,942\xspace}
\newcommand{\nqdetres}{101\xspace}
\newcommand{\nqnonres}{2,841\xspace}

\newcommand{\pukidss}{29\%\xspace}
\newcommand{\ptwom}{26\%\xspace}
\newcommand{\pgalex}{43\%\xspace}
\newcommand{\pcorr}{${\sim}$7\%\xspace}
\newcommand{\lcorr}{$-$0.039$\pm$0.064\xspace}
\newcommand{\pctwagn}{${\sim}$17\%\xspace}
\newcommand{\nages}{16,014\xspace}
\newcommand{\nexages}{14,965\xspace}
\newcommand{\ninages}{1,049\xspace}
\newcommand{\nexoff}{29\xspace}
\newcommand{\zexsig}{0.057\xspace}
\newcommand{\zinsig}{0.046\xspace}

\graphicspath{{./}{./}}
\accepted{December 4, 2020}
\submitjournal{\apj}

\shorttitle{Luminous AGNs Lacking X-ray Detections}
\shortauthors{Carroll et al.}

\begin{document}

\title{A Large Population of Luminous Active Galactic Nuclei Lacking X-ray Detections: Evidence for Heavy Obscuration?}
\author[0000-0003-3574-2963]{Christopher M. Carroll}
\affiliation{Department of Physics and Astronomy, Dartmouth College, 6127 Wilder Laboratory, Hanover, NH 03755, USA}

\author[0000-0003-1468-9526]{Ryan C. Hickox}
\affiliation{Department of Physics and Astronomy, Dartmouth College, 6127 Wilder Laboratory, Hanover, NH 03755, USA}

\author[0000-0002-7100-9366]{Alberto Masini}
\affiliation{SISSA, Via Bonomea 265, 34151 Trieste, Italy}
\affiliation{INAF - Osservatorio di Astrofisica e Scienza dello Spazio di Bologna, via Gobetti 93/3, I-40129 Bologna, Italy}
\affiliation{Department of Physics and Astronomy, Dartmouth College, 6127 Wilder Laboratory, Hanover, NH 03755, USA}

\author[0000-0002-3249-8224]{Lauranne Lanz}
\affiliation{Department of Physics and Astronomy, Dartmouth College, 6127 Wilder Laboratory, Hanover, NH 03755, USA}
\affiliation{Department of Physics, The College of New Jersey, 2000 Pennington Road, Ewing, NJ 08628, USA}

\author[0000-0002-9508-3667]{Roberto J. Assef}
\affiliation{N\'{u}cleo de Astronom\'{i}a de la Facultad de Ingenier\'{i}a, Universidad Diego Portales, Av. Ej\'{e}rcito Libertador 441, Santiago, Chile}

\author[0000-0003-2686-9241]{Daniel Stern}
\affiliation{Jet Propulsion Laboratory, California Institute of Technology, 4800 Oak Grove Drive, Mail Stop 169-221, Pasadena, CA 91109, USA}

\author[0000-0002-4945-5079]{Chien-Ting J. Chen}
\affiliation{Astrophysics Office, NASA Marshall Space Flight Center, ZP12, Huntsville, AL 35812, USA}

\author[0000-0001-8211-3807]{Tonima T. Ananna}
\affiliation{Department of Physics and Astronomy, Dartmouth College, 6127 Wilder Laboratory, Hanover, NH 03755, USA}


\begin{abstract}

We present a large sample of infrared-luminous candidate active galactic nuclei (AGNs) that lack X-ray detections in \chandra, \xmmn, and \nustar fields. We selected all optically detected SDSS sources with redshift measurements, combined additional broadband photometry from \wise, UKIDSS, 2MASS, and \galex, and modeled the spectral energy distributions (SEDs) of our sample sources. We parameterize nuclear obscuration in our SEDs with \ebv and uncover thousands of powerful obscured AGNs that lack X-ray counterparts, many of which are identified as AGN candidates based on straightforward \wise photometric criteria. Using the observed luminosity correlation between restframe \mbox{2--10 keV} (\lx) and restframe AGN 6$\mum$ (\lir), we estimate the intrinsic X-ray luminosities of our sample sources and combine these data with flux limits from X-ray catalogs to determine lower limits on nuclear obscuration. Using the ratio of intrinsic-to-observed X-ray luminosity (\rlum), we find a significant fraction of sources with column densities approaching \mbox{\nh$>$ 10$^{\text{24}}$ \cmcm}, suggesting that multiwavelength observations are necessary to account for the population of heavily obscured AGNs. We simulate the underlying \nh distribution for the X-ray non-detected sources in our sample through survival analysis, and confirm the presence of AGN activity via X-ray stacking. Our results point to a considerable population of extremely obscured AGNs undetected by current X-ray observatories.

\end{abstract}



\section{Introduction}
\label{sec:intro}

Since the first seminal study of active galactic nuclei (AGNs) by \cite{seyfert1943}, AGNs have provided astronomers with a rich field of study. From the 1950s onward, AGNs have been studied and categorized across the full electromagnetic spectrum, based on intrinsic properties such as luminosity, spectral features, and photometric colors (see \citealt{padovani2017} for review). The detection of nuclear attenuation in these systems has led to the conclusion that a considerable fraction of the AGN population is obscured by gas and dust (see \citealt{hickox2018} review and references within). This same development has provoked further investigation into how common and ``buried'' are the most extremely obscured AGNs, one constraint on which comes from modeling of the AGN contribution to the cosmic X-ray background (e.g., \citealt{ananna2019}). A commonly used tool in the identification of obscured AGNs is the presence of infrared (IR) signatures (e.g., \citealt{stern2005}), frequently obtained through photometric colors or modeling of spectral energy distributions (SEDs).

Several correlations have been observed between various properties of AGNs; in this paper we focus in particular on the relationship between X-ray and IR luminosities, commonly believed to link inverse-Compton scattering via a hot corona and reprocessed emission by the dusty torus. This relationship has been studied across low- and high-luminosity AGNs (e.g., \citealt{gandhi2009,stern2015,chen2017}), unobscured and obscured (e.g., \citealt{fiore2009,lansbury2015}), and further into the extremely obscured, Compton-thick regime (CT; i.e., \mbox{\nh $\ge$ 1.5$\times$10$^{\text{24}}$ \cmcm}) where heavy obscuration can completely attenuate the X-ray emission (e.g., \citealt{yan2019,lambrides2020}). X-ray studies of AGNs have also shown the impact different selection criteria have on the completeness and reliability of a given sample (e.g., \citealt{lamassa2019}). 

This work focuses on AGNs within X-ray survey and serendipitous fields of \chandra, \xmmn, and \mbox{\nustar} (\citealt{harrison2013}). We sought to limit our sample to straightforward observational biases (e.g., flux-limited surveys), while circumventing more complex selection effects (e.g., infrared color cuts), and no preference to selection method (e.g., X-ray selection, emission line diagnostics). We identify a population of IR-luminous AGNs lacking X-ray counterparts, despite the fact that these sources (if unobscured) should be bright enough for X-ray detection given the known correlation between X-ray and IR luminosity. We present the possibility of a population of extremely obscured AGN, a portion of which we estimate to have column densities well within the CT regime, that are currently not detected in typical X-ray observations.

This paper is organized as follows: Section \ref{sec:data} details our photometric data selection and matching to observed X-ray fields; Section \ref{sec:seds} describes our SED modeling procedure; Section \ref{sec:analysis} details our analysis and estimates of obscuring column densities; Section \ref{sec:discussion} summarizes our results and future work. Throughout the paper, we assume a $\Lambda$CDM cosmology with parameters \mbox{$H_{\text{0}}=$ 70 km s$^{-{\text{1}}}$ Mpc$^{-{\text{1}}}$}, \mbox{$\Omega_{\text{m}}=$ 0.3}, and \mbox{$\Omega_\Lambda=$ 0.7} (\citealt{spergel2007}).


\section{Data}
\label{sec:data}

\subsection{Photometry}
\label{sec:photometry}
The sample described in this paper consists of \nsrc optically detected sources by the Sloan Digital Sky Survey (SDSS; \citealt{york2000}), matched to sources detected by the \emph{Wide-field Infrared Survey Explorer} (\wise; \citealt{wright2010}), that co-exist within one or more of our X-ray datasets, consisting of the archives and survey fields of the \chandra, \xmm, and \nustar\ observatories.

We obtained all SDSS sources from Data Release 14 (\citealt{abolfathi2018}) with photometric or spectroscopic redshift measurements. SDSS provides five optical bands, $ugriz$, with median 5$\sigma$ depths for its photometric observations of 22.15, 23.13, 22.70, 22.20, and 20.71 AB mag respectively. To ensure accurate SED modeling, we restricted our sample to sources with reliable photometry (i.e., \mbox{PhotoObjAll} column \textsc{clean}=1). We used composite model magnitudes, corrected for Galactic reddening, to ensure accurate flux measurements of both point sources and extended sources. All sources in our sample have either spectroscopic or photometric redshift measurements, with the latter majority being obtained from the photometric redshift database of SDSS---the accuracy of which degrades with increasing redshift (see Figure 7 of \citealt{beck2016}). For this reason, we restricted all SDSS photometric redshifts to \mbox{$z_{\text{phot}} \le$ 0.8}, and only considered values from the Photoz table with column \textsc{photoErrorClass} of $-$1, 1, 2, or 3, corresponding to an average root-mean-square error between \mbox{0.066 $\leq$ RMSE $\leq$ 0.074}.

\begin{figure}
\epsscale{1.1}
\plotone{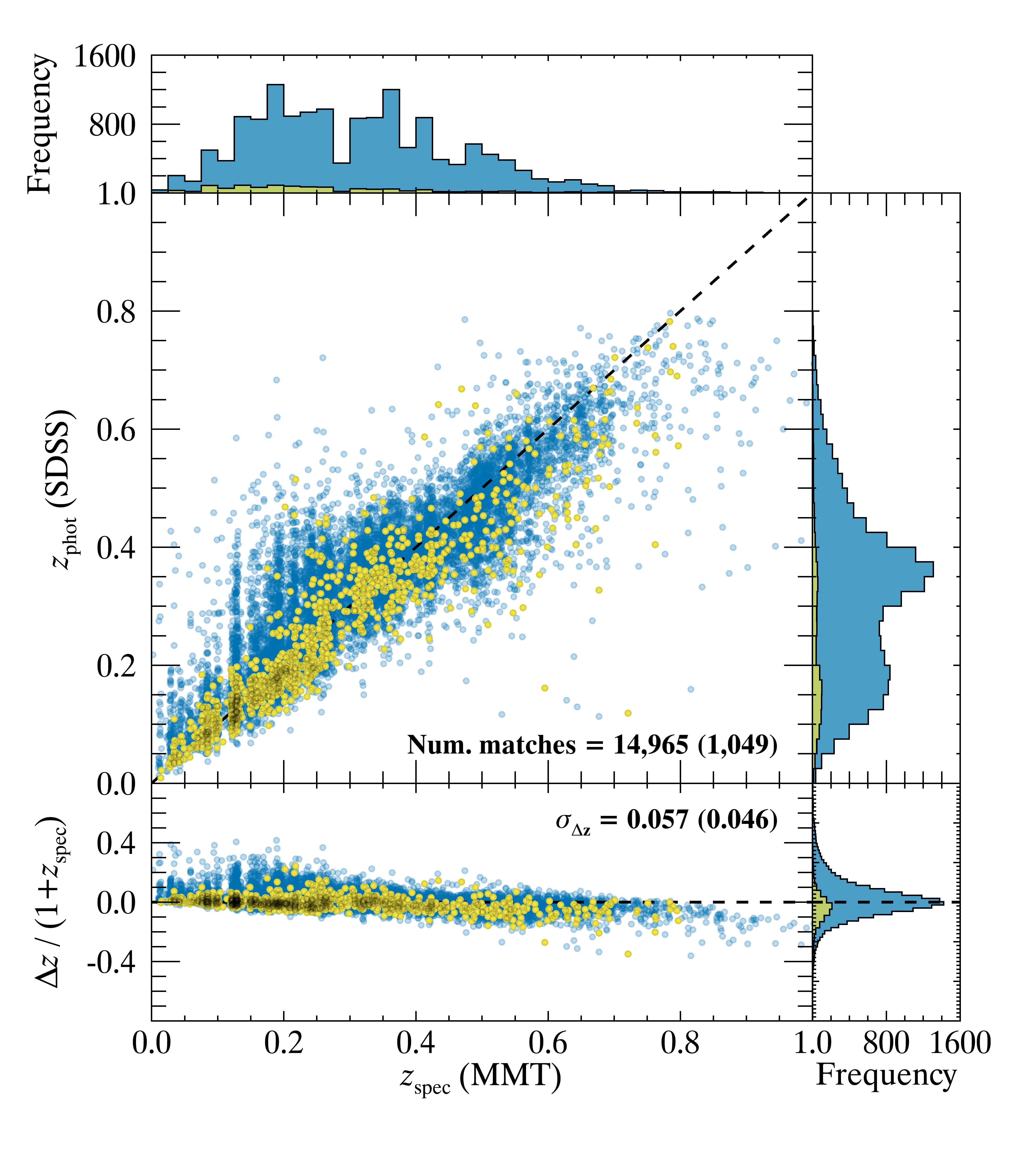}
\caption{Comparison of SDSS photometric redshifts to spectroscopic redshifts from the AGES redshift catalog. All SDSS sources were matched to the AGES catalog within 1$''$. We found a total of \nexages matches external to our sample (\emph{blue circles}), and \ninages matches to sources in our final sample (\emph{yellow circles}). Of the SDSS targets, \nexoff sources resulted in catastrophic failures when comparing their photometric redshifts to to the spectroscopically confirmed counterparts and are not explicitly shown in the figure (\mbox{$z_{\text{spec}} >$ 1}). The residuals between the SDSS and AGES redshifts are given in the lower panel, with a standard deviation of $\sigma_{\Delta z}$ = \zexsig (\zinsig for sources in our final sample). Histograms of the redshifts and residuals are also shown adjacent their respective axes.
\label{phot_spec}}
\end{figure}

The SDSS photometric redshift database was constructed using optical photometry from galaxies, and as shown in this work, some of these galaxies may contain contribution from obscured AGNs, the presence of which can have a significant impact on redshift estimation. As we rely heavily on photometric redshifts, we further compared estimates from the SDSS Photoz table to the independently measured spectroscopic redshifts from the AGN and Galaxy Evolution Survey (AGES; \citealt{kochanek2012}) in the 9 deg$^{\text{2}}$ Bo\"{o}tes field to ensure their validity for SED modeling (see Figure \ref{phot_spec}). We crossmatched all SDSS sources---with reliable photometric redshifts as described above---to the AGES catalog within 6$''$ and found \nages pairs (\nexages external to our sample) for comparison. We calculated the difference as $\Delta z / (1+z_{\text{spec}})$ and found a standard deviation of $\sigma_{\Delta z}$ = \zexsig (\zinsig for the \ninages sources matched to our sample); errors small enough that their effect on broadband SED modeling are negligible. Following this analysis, we replaced the photometric redshifts of SDSS with the spectroscopic redshifts for the AGES for sources within our sample.

\begin{figure}
\epsscale{1.1}
\plotone{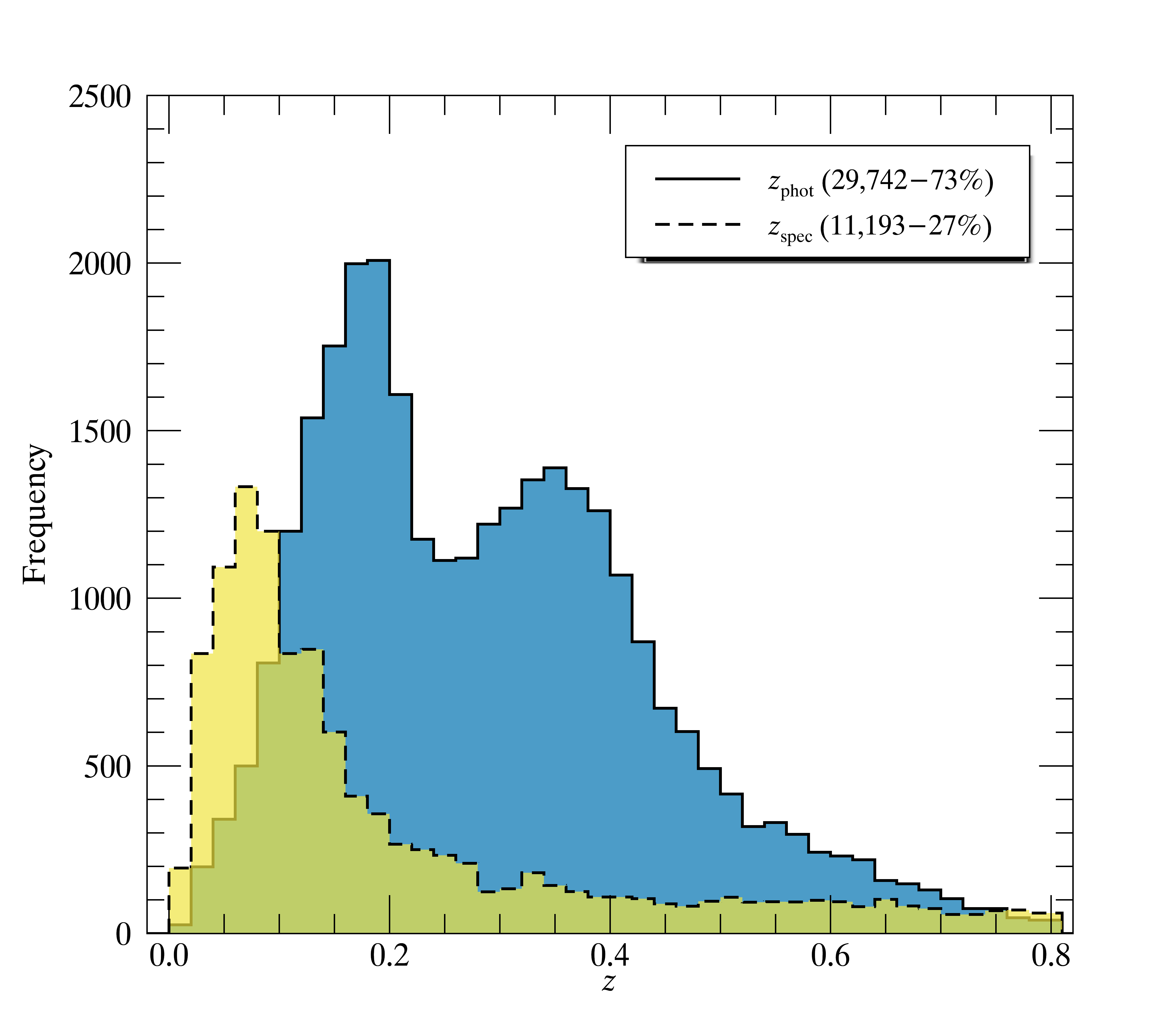}
\caption{Redshift distribution of our final sample.
\label{z_dist}}
\end{figure}

To complement our initial SDSS sample, we added additional sources with photometric redshift estimates from \xdqso (\citealt{dipompeo2015}), which primarily contains optically unobscured AGNs. We then incorporated additional spectroscopic redshift information from external surveys: \cite{reyes2008}, \cite{lacy2013}, \cite{hainline2014b}, and \cite{yuan2016}. We placed an identical constraint on all spectroscopic redshifts of \mbox{$z_{\text{spec}} \le$ 0.8} to maintain the validity of X-ray $K$-corrections. For objects with redshifts from multiple sources, we preferentially chose spectroscopic measurements from the external surveys over SDSS. Where spectroscopic redshifts were unavailable, we chose photometric estimates from \xdqso over SDSS. The redshift distribution of our final sample is shown in Figure \ref{z_dist}.

We matched our combined SDSS sources to \wise photometric data from the AllWISE Source Catalog to search for all objects within 3$''$ (half of the \wise 6$''$ resolution) of each SDSS source, pairing the AllWISE source with minimum separation for each SDSS object. The AllWISE Source Catalog provides four mid-IR (MIR) bands, with effective wavelengths of $\lambda_{\text{eff}} \sim$ 3.4, 4.6, 12.0, 22.0 $\mum$ ($\w{1}$-$\w{4}$), achieving a signal-to-noise ratio (SNR) of 5.0 at 54, 71, 730, and \mbox{5000 mJy} (16.9, 16.0, 11.5 and 8.0 Vega magnitudes), respectively. To increase the depth and maximize the SNR of our sources, we utilized ``forced photometry" from the unWISE Catalog (\citealt{lang2014}). The unWISE Catalog contains unblurred coadds of \wise imaging and photometry at known object locations using SDSS source positions; a process which helps preserve the highest SNR ratio for the \wise photometry by retaining the intrinsic resolution of the data. We replaced the MIR photometry in any band where the SNR from the unWISE Catalog is greater than that found in AllWISE. The benefits of unWISE over AllWISE are most noticeable in $\w{1}$ and $\w{2}$ where \wise is the most sensitive (see Figures 5, 6, and 7 of \citealt{lang2014}), pushing our detection threshold a few magnitudes deeper. Additionally, we required detections be present in all four \wise bands for adequate MIR coverage to constrain the AGN component of our SEDs.

We supplemented our photometric dataset with near-IR (NIR) data obtained by the UKIRT Infrared Deep Sky Survey (UKIDSS; \citealt{lawrence2007}) and the Two Micron All Sky Survey (2MASS; \citealt{skrutskie2006}). For this study, we used the UKIDSS Large Area Survey (LAS) DR10 (2013 January 14), which covers {$\sim$}4000 deg$^{\text{2}}$ of the SDSS footprint. We again matched against each of our SDSS sources within 3$''$, pairing the UKIDSS source with minimum separation distance. The UKIDSS LAS DR10 catalog provides $YJHK$ band photometry---with depths of 20.2, 19.6, 18.8, and 18.2 Vega mag---for \pukidss of the sources in our sample. We used Petrosian magnitudes to recover accurate galaxy measurements and converted these to fluxes using the standard conversion from Vega to AB magnitudes. We added additional NIR data from the 2MASS Point Source Catalog (PSC). The 2MASS PSC contains data on over 500 million sources and provides us with $JHK_{\text{s}}$ to depths of 15.8, 15.1, and 14.3 Vega mag for many \wise sources. The 2MASS photometry was only added to our sample where UKIDSS data was not available. To better match the Petrosian magnitudes of UKIDSS and standardize the NIR photometry from different sources, we utilize 2MASS 4$''$ aperture photometry. The addition of 2MASS data provided NIR data for an additional \ptwom of the sources in our initial sample.

We further supplemented our dataset with UV data obtained by the \emph{Galaxy Evolution Explorer} space mission (\galex; \citealt{martin2005}). For this study, we use the Revised All-Sky survey (GUVcat, \citealt{bianchi2017}), which covers over 22,000 deg$^{\text{2}}$ of the sky. As before, we matched to all sources within 3$''$, pairing the \galex source with minimum separation distance. The \galex GUVcat provides two ultraviolet bands, far-UV (FUV, $\lambda_{\text{eff}}\sim$ 1528 \r{A}) and near-UV (NUV, $\lambda_{\text{eff}}\sim$ 2310 \r{A})---to average depths of 19.9 and 20.8 AB mag---for \pgalex of the sources in our sample.

We followed the procedure of \cite{dipompeo2014b} and applied their angular mask to remove regions of the sky with possible IR contamination from bright stars. We further restricted all photometric data to detections with SNR $\ge$ 3.0. Although the unWISE Catalog contains sources at known SDSS positions, negating the necessity for a SNR cut, we nonetheless imposed this restriction to ensure the removal of spurious detections within the confusion limits of the unWISE processing algorithm. In all cases (except SDSS which provides deredenned magnitudes), we corrected for Galactic extinction through IR dust maps (\citealt{schlegel1998}). Finally, we required all sources in our sample to have a minimum of seven photometric bands in order to accurately model their SEDs. As a last step, we matched our final sample to the \wise AGN R90 Catalog (90\% reliability; \citealt{assef2018}) and flagged all matches. A record of our final sample and source numbers from corresponding observatories is shown in Table \ref{sample}, while the details of our data selection criteria and the effects on our sample size are shown in Table \ref{prop_cuts}.

\begin{deluxetable*}{lrrrrrrrrrr}
\tablecaption{Number of sources with various telescope coverage.\label{sample}}
\tablehead{\colhead{} & \colhead{SDSS} & \colhead{\xdqso} & \colhead{\wise} & \colhead{UKIDSS} & \colhead{2MASS} & \colhead{\galex} & \colhead{\chandra} & \colhead{\xmm} & \colhead{\nustar}}
\startdata
Initial & \nisdss & \nixdqso & \niwise & \niunwise & \nitwom & \nigalex & \nicha   & \nixmm   & \ninst   \\
Final   & \nsdss  & \nxdqso  & \nwise  & \nunwise  & \ntwom  & \ngalex  & \ninfcha & \ninfxmm & \ninfnst \\
\enddata
\tablecomments{Our initial sample of \ninit sources with optical and IR coverage, reduced to \nfinal sources for our final sample. All \wise photometry was replaced with forced photometry from unWISE where available (\niunwise initial, \nunwise final).}
\end{deluxetable*}

\begin{deluxetable*}{lrrrr}
\tablecaption{Various property selection criteria and effects on sample numbers.\label{prop_cuts}}
\tablehead{\colhead{Property Cut} & \colhead{$N_{\text{tot}}$} & \colhead{Tot. Loss} & \colhead{$N_{\text{cum}}$} & \colhead{Cum. Loss}}
\startdata
initial sample          & \ninit  & 0\%         & \ninit   & 0\%         \\
valid redshift          & \nzgood & \nzgoodtotf & \nrzgood & \nzgoodcumf \\
clean photometry        & \nclean & \ncleantotf & \nrclean & \ncleancumf \\
seven photometric bands & \nbands & \nbandstotf & \nrbands & \nbandscumf \\
four \wise bands        & \nfourw & \nfourwtotf & \nrfourw & \nfourwcumf \\
not removed by mask     & \nnomsk & \nnomsktotf & \nrnomsk & \nnomskcumf \\
not a duplicate         & \nnodup & \nnoduptotf & \nrnodup & \nnodupcumf \\
\enddata
\tablecomments{Individual and cumulative effect of each property cut on our sample. For each property cut, the first two subsequent columns list the number of sources which pass the designated cut and the fractional loss from our initial sample (\ninit sources). The last two columns track the cumulative effects of each property cut on sample size and our cumulative fractional loss.}
\end{deluxetable*}

\subsection{X-ray Data Sets and Flux Limits}
\label{sec:x-data}
The sample discussed in Section \ref{sec:photometry} is comprised of sources observed by one or more of following X-ray observatories: \chandra, \xmm, or \nustar. In this work, we utilize five X-ray survey data sets: the \chandra Source Catalog 2 (CSC2; \citealt{evans2010}), the 3XMM-DR8 catalogue (16 May 2018; \citealt{rosen2016}), and the combined \nustar catalogs of COSMOS and UDS (\citealt{civano2015,masini2018}), as well as the \nustar serendipitous survey (SSC; \citealt{lansbury2017}). For an X-ray observation to be considered, the specific observation must be included in one of the five X-ray survey data sets used in this work. For a source to be considered ``in-field'' (covered by an X-ray observation), it must reside within the FOV of an observation ($16.9'\times16.9'$ for \chandra, $33'\times33'$ for \xmm, $13'\times13'$ for \nustar). We selected only \chandra fields observed with ACIS-I and used only \xmm fields with full-frame observations (\textsc{pn\_mode} set to \textsc{flg}, \textsc{ff}, or \textsc{eff}).

From CSC2, we removed observations where any of the following flags are present, unless manually added after consideration (i.e., \textsc{man\_add\_flag} is \textsc{true}): \textsc{dither\_warning\_flag}, \textsc{pileup\_flag}, \textsc{sat\_src\_flag}, \textsc{var\_flag}, \textsc{streak\_src\_flag}, \textsc{var\_inter\_hard\_flag}). From CSC2, we use aperture model energy fluxes given by the PSF 90\% enclosed counts fraction, inferred from the canonical absorbed power law model, corrected to the full PSF (\textsc{flux\_powlaw\_aper{\small90}}$*1.1$). For this study, we focused primarily on observations with the ACIS detector to coincide with ACIS-I observations from \chanmaster. After applying these data criteria, we found \ninfcha sources from our sample exist within the \chandra footprint.

From 3XMM, we included only sources without the chance for spurious detections (\textsc{sum\_flag} of 0 or 1). We included only observations with full CCD chip readout (\textsc{pn\_submode} of \textsc{pfw} or \textsc{pfwe}) and focused only on detections with the EPIC-PN detector, which has a higher quantum efficiency at low and high energies compared to the dual MOS chips. After applying this data criteria, we found \ninfxmm sources exist within the \xmm footprint.

Our \nustar dataset is the combination of multiple catalogs, each of which are treated individually. For sources without soft X-ray counterparts, we used either the \nustar positions (COSMOS) or optical counterpart positions (SSC). We removed sources where the data prohibit reliable photometric constraints (SSC; see Section A.6 of \citealt{lansbury2017}). We removed upper limits from all catalogs and took the average of the net (A+B) exposure time in seconds where necessary (SSC and UDS). After applying this data criteria, we found \ninfnst sources exist within the \nustar footprint.

We then matched all in-field sources to our X-ray data sets using a maximum separation distance of 6.25$''$ for all three X-ray observatories, adopting this value from the \xmm PSF FWHM of 12.5$''$ to account for positional offsets. We then calculated observed X-ray luminosities (\lx), converting the fluxes of all detections to the \mbox{2--10 keV} energy band via the online tool WebPIMMS\footnote{\url{https://heasarc.gsfc.nasa.gov/cgi-bin/Tools/w3pimms/w3pimms.pl}}, assuming an average Galactic \mbox{\nh = 2$\times$10$^{\text{20}}$ \cmcm} and a photon index of \mbox{$\Gamma$ = 1.8}. We $K$-corrected all fluxes to the rest frame, using the redshift of each source from our main sample. For sources with X-ray detections in multiple energy bands, we preferentially selected detections which required the least fractional energy conversion factor to reach restframe \mbox{2--10 keV}.

\begin{figure*}
\epsscale{1.1}
\plotone{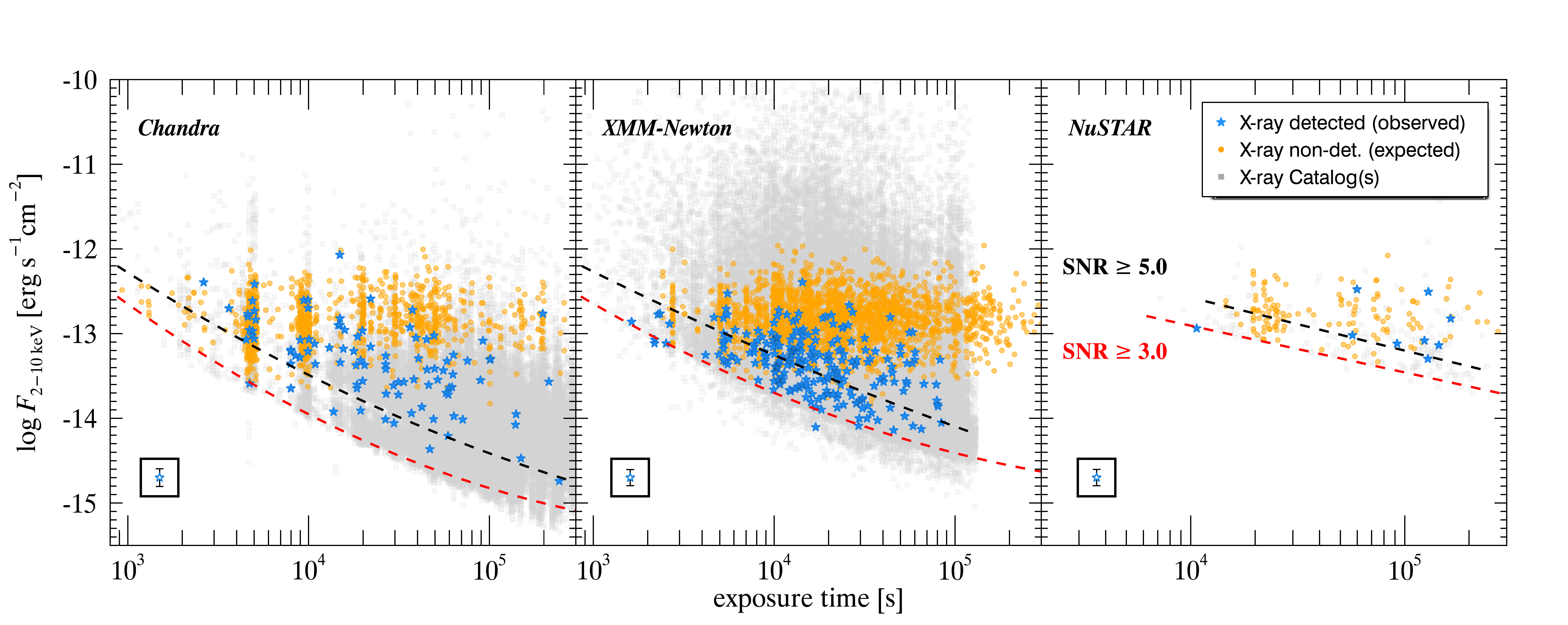}
\caption{X-ray 2--10 keV flux vs. exposure time for the three observatories in our sample. We show X-ray catalog (\emph{grey squares}) and SED-selected AGN sample (\emph{blue stars}) sources alongside intrinsic X-ray flux estimates (\emph{orange circles}) for our non-detected sources. The X-ray flux limits were inferred from the \lx--\lir relationship (see Section \ref{sec:lum_ratio}), with exposure times obtained from the master observations catalog of each instrument. We estimated X-ray flux limits calculated from catalog sources with SNR $\ge$ 3.0 (\emph{red dashed line}) and SNR $\ge$ 5.0 (\emph{black dashed line}). The mean error is depicted in the lower left of each panel.
\label{fxlim}}
\end{figure*}

The remaining, unmatched catalog sources from each data set were then used to estimate flux limits of each observatory, for both \mbox{SNR $\ge$ 3.0} and \mbox{SNR $\ge$ 5.0} (see Figure \ref{fxlim}).

\subsection{X-ray Non-Detections}
\label{sec:non-det}

For sources without X-ray counterparts, we used the master observations catalog of each X-ray telescope to estimate upper limit X-ray fluxes. Each respective master catalog provides exposure times for all observed fields, and we crossmatched each catalog source to a master field and determined the distance from field center. In this work, we did not directly account for vignetting, and instead chose to limit our scope to sources where the reduction in effective area due to vignetting was less than 30\% (FOV$_{\text{\emph{eff}}}$; 7$'$ for \chandra and \xmm, and 5$'$ for \nustar). We adopted the exposure times of each corresponding master field for our X-ray non-detected sources. In cases where a source exists within multiple field observations of the same instrument, we coadded all exposure times---within the selected FOV$_{\text{\emph{eff}}}$---as is customary for the respective survey catalogs. Using the assigned exposure times, we estimated the upper limit X-ray flux per source and calculated upper limit X-ray luminosities (\lxlim) for all X-ray non-detected sources.

SDSS source positions and X-ray luminosity estimates are shown in Table \ref{xlum}.

\begin{deluxetable*}{rrrrrrrrr}
\tablecolumns{9}
\tablecaption{Source positions and X-ray luminosity estimates by observatory.\label{xlum}}
\tablehead{\colhead{SDSS ObjID} & \colhead{RA} & \colhead{Dec} & \colhead{$\log L_{\text{X}}(\chandra)$} & \colhead{$\log L_{\text{X}}(\xmm)$} & \colhead{$\log L_{\text{X}}(\nustar)$} & \colhead{} \\
\colhead{} & \colhead{[deg]} & \colhead{[deg]} & \colhead{[erg\,s$^{-1}$]} & \colhead{[erg\,s$^{-1}$]} & \colhead{[erg\,s$^{-1}$]} & \colhead{}}
\startdata
1237667223933550815 &   4.7399 & -20.3684 &           \nodata &    43.11$\pm$0.09 &           \nodata &     XMM \\
1237667223399170264 &  10.7800 & -20.6891 & $<$41.25$\pm$0.16 & $<$41.81$\pm$0.16 &           \nodata &     CHA \\
1237667223399170855 &  10.8610 & -20.5961 & $<$42.46$\pm$0.15 & $<$43.03$\pm$0.15 &           \nodata &     CHA \\
1237667223399170348 &  10.7717 & -20.6602 &           \nodata &    44.28$\pm$0.09 &           \nodata &     XMM \\
1237667223387439337 & 342.3175 & -19.3099 &           \nodata & $<$41.47$\pm$0.15 &           \nodata &     XMM \\
1237667223387504664 & 342.4005 & -19.2649 &           \nodata & $<$40.37$\pm$0.21 &           \nodata &     XMM \\
1237671166167941234 & 166.6228 & -18.3672 &           \nodata &    44.01$\pm$0.06 &           \nodata &     XMM \\
1237667225535119631 & 343.0232 & -17.7923 &           \nodata & $<$41.42$\pm$0.17 &           \nodata &     XMM \\
1237667223389471530 & 347.1830 & -19.8406 &           \nodata & $<$43.05$\pm$0.15 &           \nodata &     XMM \\
1237667225544556806 &   5.6148 & -18.9625 &           \nodata & $<$42.67$\pm$0.15 & $<$43.06$\pm$0.15 &     XMM \\
\enddata
\tablecomments{Upper limit X-ray luminosity estimates are labeled by ``$<$'. The field chosen for our analysis is shown in the final column.}
\end{deluxetable*}


\section{SED Modeling}
\label{sec:seds}
We modeled our SEDs following the methodology described in \citet[hereafter A10]{assef2010}, fitting our photometry with a non-negative linear combination of galaxy and AGN templates. The empirically derived models of A10 consist of three galaxy templates representing different stellar populations (passive, star-forming, and starburst) and a ``typical'' unobscured AGN template. A coadded multi-galaxy template approach allows us to fit SEDs with multiple stellar populations and mitigate the overestimation of AGN contribution. In order to be as conservative as possible in assigning AGN contribution at MIR wavelengths where SEDs are constrained by at most four photometric data points, we replaced the star-forming template of A10 with a star-forming galaxy (SFG) template from \citet[hereafter K15]{kirkpatrick2015}. The star-forming and starburst templates of A10 are nearly identical at wavelengths longer than 6 micron, while on average, the SFG templates of K15 contribute more MIR flux than that of A10. Of the three star-forming templates of K15, we have chosen SFG1 as it displays a minimum redshift evolution in $\w{1}-\w{2}$ color (i.e., the \wise color space most sensitive to increasing obscuration). We flux-normalized the SFG1 template and replaced the star-forming template of A10 redward of 2 micron, where the K15 templates begin.

To model nuclear obscuration, we followed A10 and applied a reddening law to the AGN component which consists of a Small Magellanic Cloud (SMC) like extinction curve for \mbox{$\lambda <$ 3300 \r{A}} (\citealt{gordon1998}) and a Galactic extinction curve at longer wavelengths (\citealt{cardelli1989}), assuming a ratio of total-to-selective extinction of \mbox{$R_{V}$ = 3.1} (see A10 for details). We parameterized the AGN reddening as \ebv and simulate levels of obscuration in the range of \mbox{0.0 $\le$ \ebv $\le$ 50.0} in bins of \mbox{$\Delta$\ebv = 0.05 dex} (offset to 0), and spanning a redshift range of \mbox{0.00 $\le z <$ 1.00} in bins of \mbox{$\Delta z$ = 0.001}. We convolve the templates individually with the instrument response curves for each bandpass and construct a grid based parameter space for each of our templates which we coadded to match our photometry.

\begin{figure}
\epsscale{1.2}
\plotone{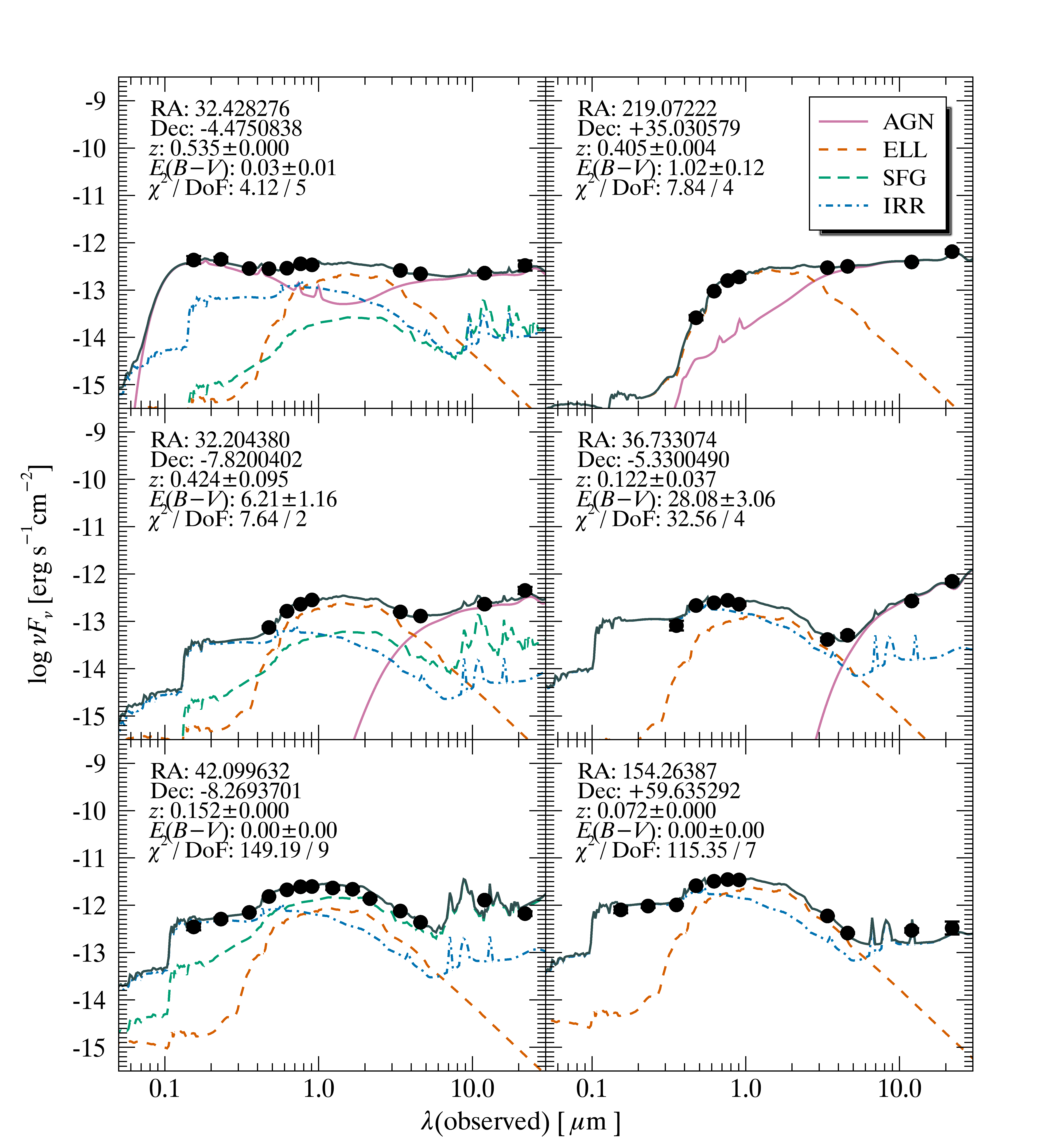}
\caption{Example SEDs. Each SED is comprised of a combination of four template components: AGN (\emph{solid, magenta}), elliptical/passive (\emph{dashed, orange}), star-forming (\emph{long dash, green}), and irregular/starburst (\emph{dash-dot, blue}). The AGN component includes attenuation to simulate nuclear obscuration. The top four SEDs in this figure are all confirmed \wise AGNs; the top panel sources have X-ray counterparts, while the middle panel sources do not. The bottom panel sources are SEDs with no AGN contribution and are shown for comparison.
\label{seds}}
\end{figure}

We then modeled the SEDs of our sources to identify the presence of AGN activity. A selection of our best-fit SED models is shown in Figure \ref{seds}. For each source, we modeled the full parameter space and performed \chisq minimization to determine the contribution from the AGN and three galaxy templates (component normalizations $C_{\text{AGN}}$, $C_{\text{ELL}}$, $C_{\text{SFG}}$, and $C_{\text{IRR}}$) for each combination of redshift and extinction. The template normalizations are linear with respect to the \chisq statistics, and we used this to our advantage by applying a linear least squares approach, which is less computationally expensive than other, more sophisticated modeling methods. We found an exact solution for each redshift-extinction pair for all possible combinations of templates. Where our routine identified a negative template normalization---galaxy or AGN---it was set to zero before we carried out \chisq minimization. The SED with minimum reduced \chisq was chosen as our best-fit model. The best-fit \ebv and template normalizations were recorded for each source. For each best-fit SED, we calculated restframe AGN fraction ($f_{\text{AGN}}$), given as the fraction of the absorbed AGN flux to the total coadded flux at 15$\mum$.

To estimate errors on the parameters, we created 1,000 realizations of each source and resampled the redshift and photometric measurements within their uncertainties. We then chose the median \ebv and selectied the realization closest to the median. Where a source had multiple realizations with the same \ebv closest to the median, we chose the realization with minimum reduced \chisq. For each source we recorded the percent of resampled SED realizations which contained AGN contribution ($P_{\text{AGN}}$; i.e., AGN component normalization $C_{\text{AGN}} >$ 0).

From our SED models, we calculated AGN MIR luminosities (\lir) by interpolating the deredenned AGN component flux of each source at 6$\mum$. For some sources (\pcorr of sources with an AGN SED component contribution), the AGN template of A10---which is a single template consisting of a combined accretion disk and torus---is insufficient to model the data (i.e., the optical data is well fit at the expense of the IR data). This is not surprising as a single template cannot account for differences in the geometry or variability of a given source, nor does our modeling account for AGNs bluer in the UV/optical than the base AGN template. In these cases, we adjusted \lir by the difference between the total coadded model and the interpolated 6-micron flux of the source, which we interpolated from the photometry. The mean applied luminosity correction factor was \mbox{$\log L_{\text{corr}} =$ \lcorr}. It is worth noting that the redshift distribution of luminosity corrected sources follows Fig. \ref{z_dist}, with redshifts at preferentially lower values (\mbox{$z <$ 0.2}). Approximately 80\% of the cases where the AGN template failed to sufficiently model both optical and IR data were unobscured to moderately obscured (\mbox{\ebv $\leq$ 1.0}), suggesting that the uncertainties on the optical data more heavily constrained the overall SED than did the IR data. The output of our SED modeling is detailed in Table \ref{sed_output}.

\begin{deluxetable*}{rrrrrrrrr}
\tablecolumns{9}
\tablecaption{SED modeling output parameters.\label{sed_output}}
\tablehead{\colhead{$z$} & \colhead{$E(\bv)_{\text{AGN}}$} & \colhead{$C_{\text{AGN}}$} & \colhead{$C_{\text{ELL}}$} & \colhead{$C_{\text{SFG}}$} & \colhead{$C_{\text{IRR}}$} & \colhead{$\chi^2 /$ DoF} & \colhead{$P_{\text{AGN}}$} & \colhead{$\log L_{\text{MIR}}$} \\
\colhead{} & \colhead{} & \colhead{} & \colhead{} & \colhead{} & \colhead{} & \colhead{} & \colhead{[\%]} & \colhead{[erg\,s$^{-1}$]}}
\startdata
0.348$\pm$0.039 &  6.21$\pm$2.91 & 5.04$\times$10$^{-17}$ & 7.22$\times$10$^{-16}$ & 0.00$\times$10$^{+00}$ & 1.17$\times$10$^{-16}$ &  14.61 / 4 & 100 & 44.05$\pm$0.07 \\ 
0.112$\pm$0.030 & 50.00$\pm$0.50 & 5.32$\times$10$^{-17}$ & 1.22$\times$10$^{-15}$ & 0.00$\times$10$^{+00}$ & 1.53$\times$10$^{-16}$ &  27.09 / 8 &  86 & 43.06$\pm$0.01 \\ 
0.391$\pm$0.060 &  6.98$\pm$2.61 & 6.14$\times$10$^{-17}$ & 0.00$\times$10$^{+00}$ & 9.47$\times$10$^{-16}$ & 0.00$\times$10$^{+00}$ &  38.95 / 5 & 100 & 44.24$\pm$0.10 \\ 
0.778$\pm$0.172 &  0.06$\pm$0.00 & 7.05$\times$10$^{-17}$ & 3.93$\times$10$^{-16}$ & 0.00$\times$10$^{+00}$ & 0.00$\times$10$^{+00}$ &  30.68 / 6 & 100 & 44.92$\pm$0.13 \\ 
0.180$\pm$0.030 & 35.38$\pm$4.33 & 7.49$\times$10$^{-17}$ & 9.09$\times$10$^{-16}$ & 0.00$\times$10$^{+00}$ & 1.38$\times$10$^{-16}$ &  19.44 / 6 &  97 & 43.63$\pm$0.01 \\ 
0.055$\pm$0.017 & 25.02$\pm$3.06 & 1.86$\times$10$^{-16}$ & 3.19$\times$10$^{-15}$ & 7.72$\times$10$^{-16}$ & 5.65$\times$10$^{-16}$ &  53.70 / 6 & 100 & 42.98$\pm$0.05 \\ 
0.555$\pm$0.088 &  0.04$\pm$0.00 & 9.47$\times$10$^{-17}$ & 2.65$\times$10$^{-16}$ & 0.00$\times$10$^{+00}$ & 5.04$\times$10$^{-17}$ &  57.13 / 6 & 100 & 44.75$\pm$0.10 \\ 
0.096$\pm$0.034 & 50.00$\pm$0.50 & 9.97$\times$10$^{-17}$ & 6.61$\times$10$^{-16}$ & 0.00$\times$10$^{+00}$ & 1.69$\times$10$^{-16}$ &   9.59 / 5 & 100 & 43.20$\pm$0.01 \\ 
0.367$\pm$0.051 & 50.00$\pm$0.50 & 1.25$\times$10$^{-16}$ & 3.68$\times$10$^{-16}$ & 0.00$\times$10$^{+00}$ & 2.31$\times$10$^{-17}$ &  26.89 / 3 &  97 & 44.50$\pm$0.01 \\ 
0.328$\pm$0.115 & 39.71$\pm$5.45 & 7.65$\times$10$^{-17}$ & 1.79$\times$10$^{-16}$ & 0.00$\times$10$^{+00}$ & 2.71$\times$10$^{-17}$ &   1.56 / 3 &  98 & 44.18$\pm$0.02 \\ 
\enddata
\tablecomments{Modeling output parameters redshift, color excess \ebv, template normalizations (AGN, elliptical, star-forming, irregular/starburst), chi-squared per degrees of freedom, percentage of resampled realizations with AGN contribution ($C_{\text{AGN}} > 0$), and AGN $6\mum$ luminosity. Uncertainties on \ebv and $\log L_{\text{IR}}$ were estimated via the median absolute deviation of all source realizations containing AGN contribution.}
\end{deluxetable*}


\section{Analysis and Results}
\label{sec:analysis}

\subsection{Quality Cuts and Analysis Subset}
\label{sec:subset}

Following our SED modeling, additional quality cuts were applied to our final sample to ensure the accuracy of our analysis. In order to focus on objects that are robustly identified as AGN from the photometric fitting, we restricted our final sample to sources where SED models that exhibited a high AGN fraction (\mbox{$f_{\text{AGN}} \ge$ 0.7}) at 15$\mum$, assuming the majority of emission at this wavelength is from AGN contribution (e.g., \citealt{lambrides2020}). We further limited our analysis to IR-luminous sources (\mbox{\lir $\ge$ 10$^{\text{42.0}}$ erg s$^{-{\text{1}}}$}). As the SED templates used in our modeling procedure were empirically derived averages over many thousands of sources, they are not mutable enough to account for variation of any individual object in our sample. In this instance, the \chisq statistic is less meaningful for quantitative purposes and more for qualitative goodness of fit testing. We assigned a discretionary reduced chi-square cut to our SED models to remove overtly poor fits (\chisqred $\le$ 20.0).

To ensure reliable X-ray flux measurements, we only accepted X-ray detections with \mbox{$\text{SNR} \ge$ 3.0}, and only selected sources within the FOV$_{\text{\emph{eff}}}$ detailed in Section \ref{sec:non-det}. To ensure a straightforward comparison between X-ray and non X-ray detected WISE AGNs, only X-ray non-detected sources that would have exceeded the X-ray flux limits after corrected for obscuration were added to our final sample (see Figure \ref{fxlim}). Sample sources where matched X-ray data was removed due to insufficient quality were also removed from further analysis.

The combined effects of our SED and X-ray quality cuts are detailed in the ``subset'' column of Table \ref{qual_cut}. This subset consists of \nqdet X-ray detected sources and \nqnon X-ray non-detected sources. For the remainder of this paper, we refer only to these \nqagn sources in the analysis subset of our final sample.

\begin{deluxetable}{llrr}
\tablecolumns{3}
\tablecaption{Number of sources with various selections.\label{qual_cut}}
\tablehead{\colhead{Selection} & & \colhead{$N_{\text{final}}$} & \colhead{$N_{\text{subset}}$}}
\startdata
\multirow{4}{*}{Final sample} & SED Galaxy     & \ngal    & \nqgal    \\
                              & SED Galaxy+AGN & \nagn    & \nqagn    \\
                              & X-ray detected & \ndet    & \nqdet    \\
                              & X-ray non-det. & \nnon    & \nqnon    \\
                              \cline{2-4}
\multirow{3}{*}{\wise AGN}    & in catalog     & \ninfwac & \nqinfwac \\
                              & X-ray detected & \ndetwac & \nqdetwac \\
                              & X-ray non-det. & \nnonwac & \nqnonwac \\
                              \cline{2-4}
\multirow{3}{*}{\chandra}     & in X-ray field & \ninfcha & \nqinfcha \\
                              & X-ray detected & \ndetcha & \nqdetcha \\
                              & X-ray non-det. & \nnoncha & \nqnoncha \\
                              \cline{2-4}
\multirow{3}{*}{\xmm}         & in X-ray field & \ninfxmm & \nqinfxmm \\
                              & X-ray detected & \ndetxmm & \nqdetxmm \\
                              & X-ray non-det. & \nnonxmm & \nqnonxmm \\
                              \cline{2-4}
\multirow{3}{*}{\nustar}      & in X-ray field & \ninfnst & \nqinfnst \\
                              & X-ray detected & \ndetnst & \nqdetnst \\
                              & X-ray non-det. & \nnonnst & \nqnonnst \\
\enddata
\tablecomments{Our final sample of \nsrc sources, reduced to \nqagn sources for our analysis subset.}
\end{deluxetable}

\subsection{X-ray Luminosity Ratio}
\label{sec:lum_ratio}
\begin{figure}
\epsscale{1.1}
\plotone{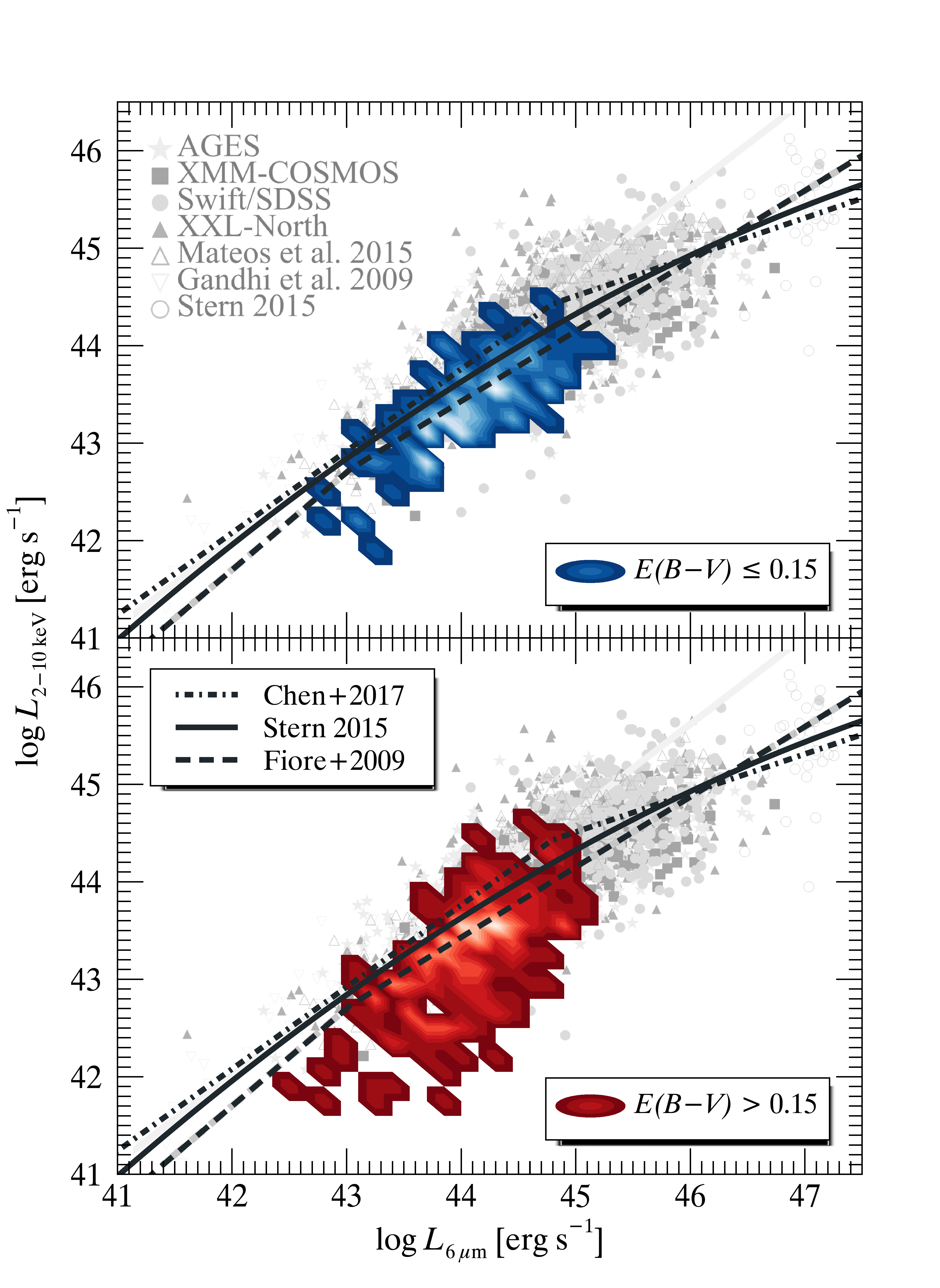}
\caption{Observed X-ray luminosity (\lx) versus the AGN MIR luminosity (\lir), compared to observations for unobscured quasars (taken from C17) as well as best-fit relationships for unobscured AGN from C17 and F09. We separate our sample into two categories, SED unobscured sources (\mbox{\ebv $\le$ 0.15}; \emph{upper panel, blue contours}) and obscured sources (\mbox{\ebv $>$ 0.15}; \emph{lower panel, red contours}), and show their sample density overlaid on Figure 2 of C17 (reprinted with permission).
\label{lx_lir}}
\end{figure}

We compared the calculated X-ray and MIR luminosities of our X-ray detected sources to the X-ray-MIR luminosity relationships for unobscured AGN presented by \citet[hereafter F09]{fiore2009}, \citet[hereafter S15]{stern2015}, and \citet[hereafter C17]{chen2017}. Figure \ref{lx_lir} illustrates that our SED unobscured sample is generally consistent with the dataset of C17 in the relationship between the X-ray and IR luminosities of unobscured AGNs. We interpreted this as further confirmation of the validity of our SED modeling and AGN MIR luminosities. Though the X-ray and IR luminosities for our unobscured sources are broadly consistent with the C17 relationship, our sample lies systematically below the \lx--\lir relation of C17. Extensive testing has shown that the systematic offset (\mbox{{$\sim$}0.3 dex}) between the C17 dataset and our sample is due to differences in our target selection methods and SED modeling procedures. A parallel analysis was performed with all three \lx--\lir relations and we found no qualitative difference in our results. In what follows, we have adopted the \lx--\lir relation of C17 as it is the most robust study of the known relationship to date, and use this relationship for the rest of our analysis. Using the \lx--\lir relation, we estimated intrinsic X-ray luminosities as a function of AGN MIR luminosity (\lxir) for the sources in our sample.

\begin{figure*}
\epsscale{0.8}
\plotone{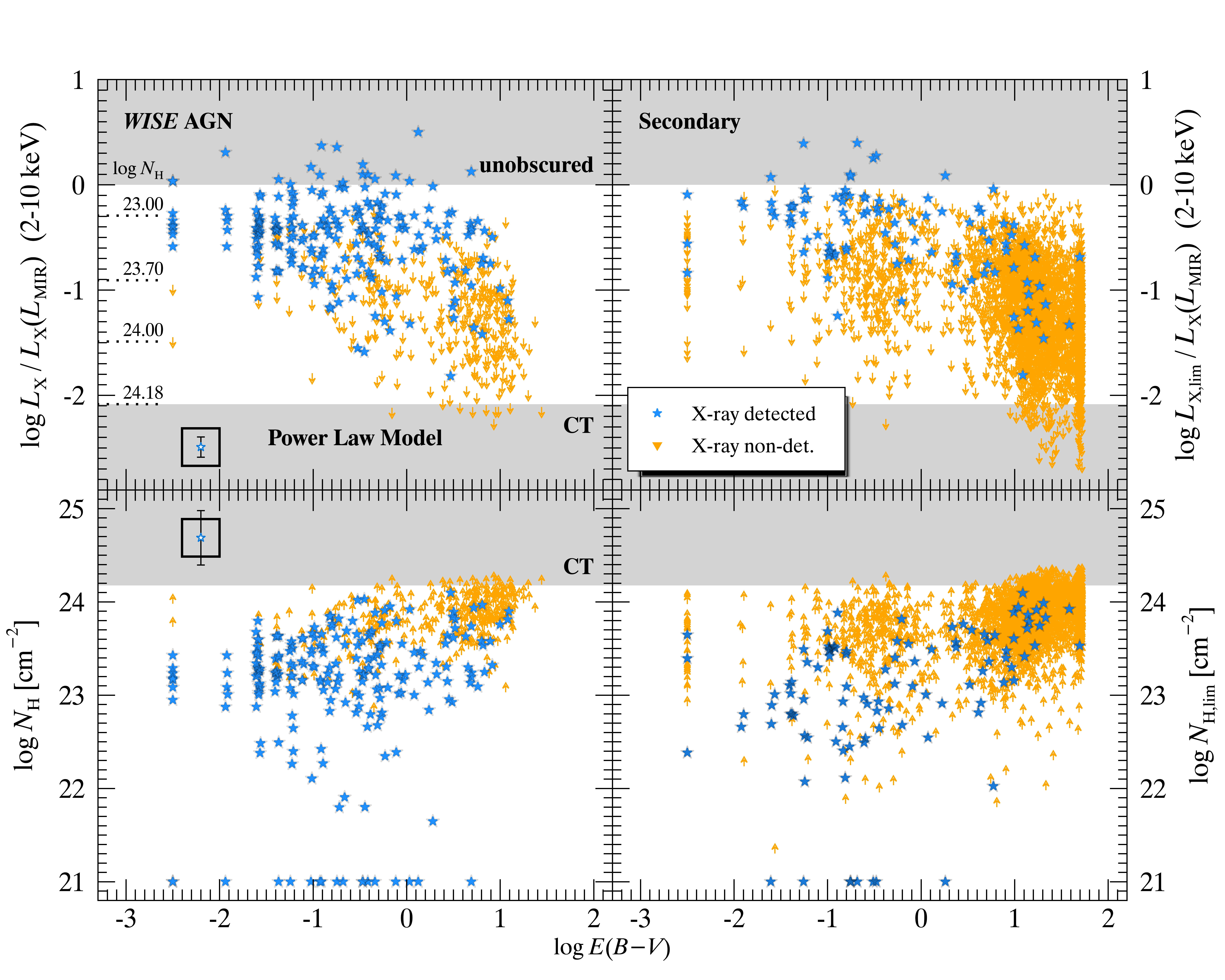}
\plotone{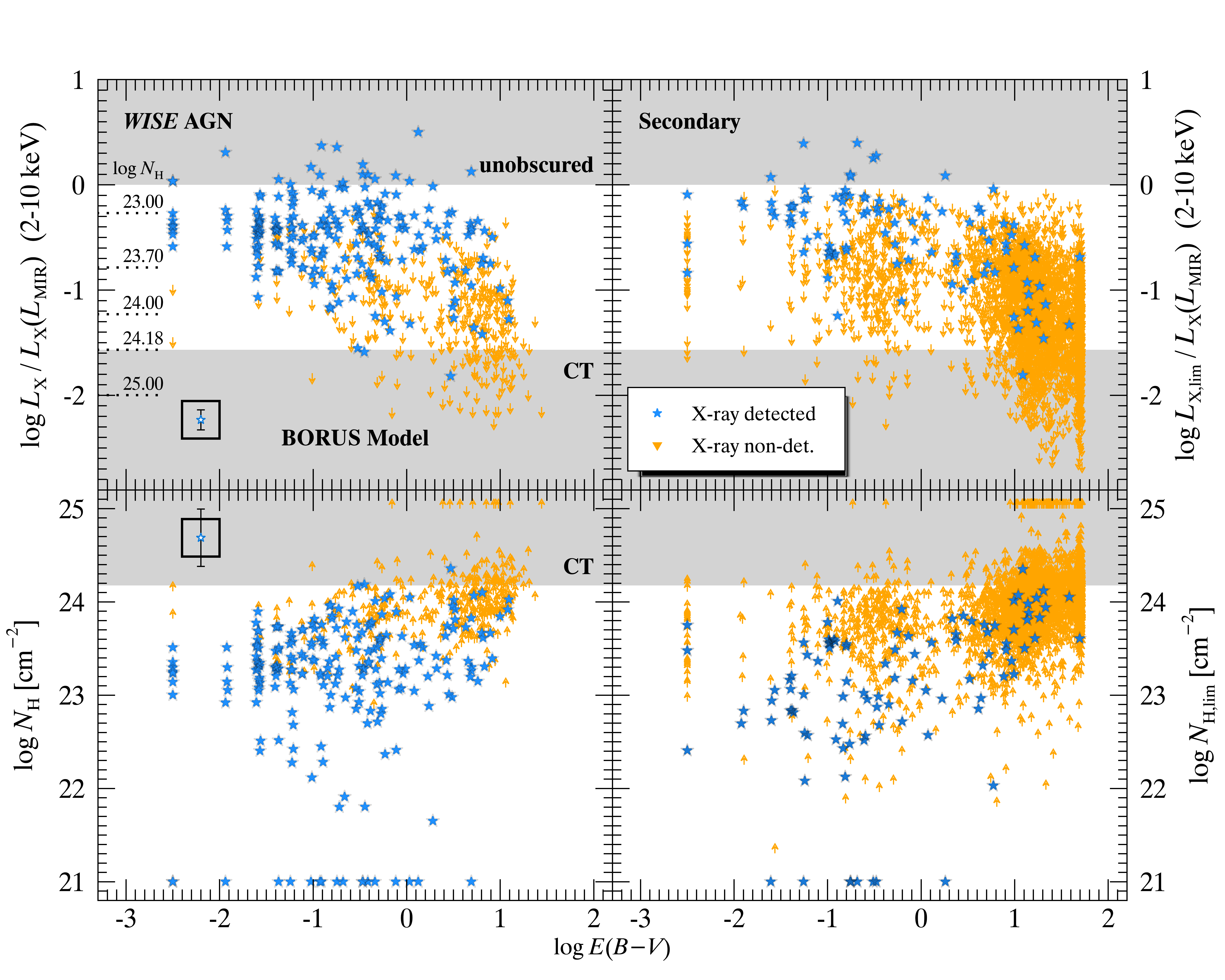}
\caption{Luminosity ratio \rlum and \nh estimates for the power law and BORUS X-ray spectral models. In each figure, the top panels show \rlum as a function of nuclear obscuration \ebv, while the bottom panels display the corresponding \nh estimates inferred from \rlum. X-ray detected (\emph{blue stars}) and non-detected (\emph{orange arrows}) sources are further grouped with \wise AGNs in the left columns and secondary sources in the right columns. In each panel, the grey shaded areas highlight unobscured and CT obscuration (labelled accordingly), assuming the given X-ray spectral model. For X-ray detected sources, \rlum is the ratio of \lx$/$\lxir. For X-ray non-detected sources (\emph{orange arrows}), \rlum is the ratio of \llim$/$\lxir, and are considered upper limits, and corresponding \nh values lower limits.
\label{lratio}}
\end{figure*}

With X-ray luminosities and our intrinsic X-ray luminosity estimates, we calculated the ratio of observed-to-intrinsic X-ray luminosity (\rlum) for our X-ray detected sources, which can be used to estimate the amount of obscuring material between an AGN and observer. As the \lx--\lir relation is drawn from unobscured AGN samples, \lxir can be used to provide an estimates of intrinsic X-ray luminosities even for obscured AGNs. For each X-ray non-detected source, we assigned the flux limit of Figure \ref{fxlim} in lieu of an X-ray observation and computed an upper limit \rlum based on the 3$\sigma$ upper limit on the X-ray luminosity. In Figure \ref{lratio}, we show \rlum against the AGN color excess parameter \ebv of our models. For X-ray detected sources, \rlum is simply the ratio of observed-to-intrinsic X-ray luminosity, \lx/\lxir; for X-ray non-detected sources, \rlum is the ratio of \llim$/$\lxir. Figure \ref{lratio} shows a trend between \rlum and \ebv (Kendall rank correlation \mbox{$\tau$ = $-$0.408} for \wise AGNs and \mbox{$\tau$ = $-$0.293} for secondary sources), with the majority of X-ray detected sources displaying less obscuration (\mbox{log \ebv $<$ 0.0}; \mbox{\rlum $> -$1.0}) than the X-ray non-detected sources (\mbox{log \ebv $\ge$ 0.0}; \mbox{\rlum $< -$1.0}). As \ebv represents nuclear obscuration, we would expect to see a decrease in \rlum with rising \ebv, as the observed X-ray emission is increasingly buried under gas and dust.

\subsection{Estimates of Obscuring Column Density}
\label{nh_est}
X-ray spectral modeling can be used to infer the geometry and physical properties of AGNs, including the amount of obscuring material between an observer and the galactic nucleus. Using XSPEC (v.12.10.1; \citealt{arnaud1996}), we produced model AGN X-ray spectra using two different input models: a simple power law representing the most simplistic, base AGN model, and a more sophisticated model which includes an AGN torus using the Monte Carlo radiative transfer code \borus (\citealt{balokovic2018}). For the simplistic AGN model, we chose a single power law with photoelectric absorption and optically-thin Compton scattering. For the more sophisticated \borus model, we chose a uniform density sphere with polar cutouts and a cutoff power law intrinsic continuum (borus02\_v170323a.fits\footnote{\url{http://www.astro.caltech.edu/~mislavb/download}}), with the torus covering factor and cosine of inclination as tunable parameters. For both models, we increased the line-of-sight column density from unobscured to obscured (\mbox{21.0 $\le$ log \nh/\cmcm $\le$ 25.0}, in steps of 0.25) and calculated model \mbox{2--10 keV} fluxes as the density of obscuring material increased. We converted these fluxes to luminosities, and calculated \rlum at each step assuming an intrinsic X-ray luminosity based on the unobscured model luminosity at \mbox{log \nh/\cmcm = 21.0}. Such spectral modeling of AGNs allows us to associate absorption of observed X-ray emission \rlum to nuclear line-of-sight \nh obscuration. As the values of \rlum for non-detected sources are derived from flux limits, the calculated \nh for these sources are, in turn, lower limits. We then interpolated the value of \rlum for each source to estimate the column density of nuclear obscuring material \nh. As shown in the bottom panels of Figure \ref{lratio}, many of the X-ray non-detected sources show high levels of obscuration, with some reaching into the CT regime. In fact, over half of all confirmed \wise AGNs in our analysis subset are X-ray non-detected, with roughly \pctwagn of these sources displaying lower limits already estimated to be within the CT regime assuming the \borus model. As \nh estimates for X-ray non-detected sources are all lower limits, there may be a considerable portion of the X-ray non-detected sources pushing the CT limit. The choice of X-ray spectral model and the effect on our results can also be seen in the bottom panels of Figure \ref{lratio}; using the more sophisticated and realistic BORUS model yields a higher percentage of CT sources. This result suggests a population of extremely obscured AGNs.

\begin{figure}
\epsscale{1.1}
\plotone{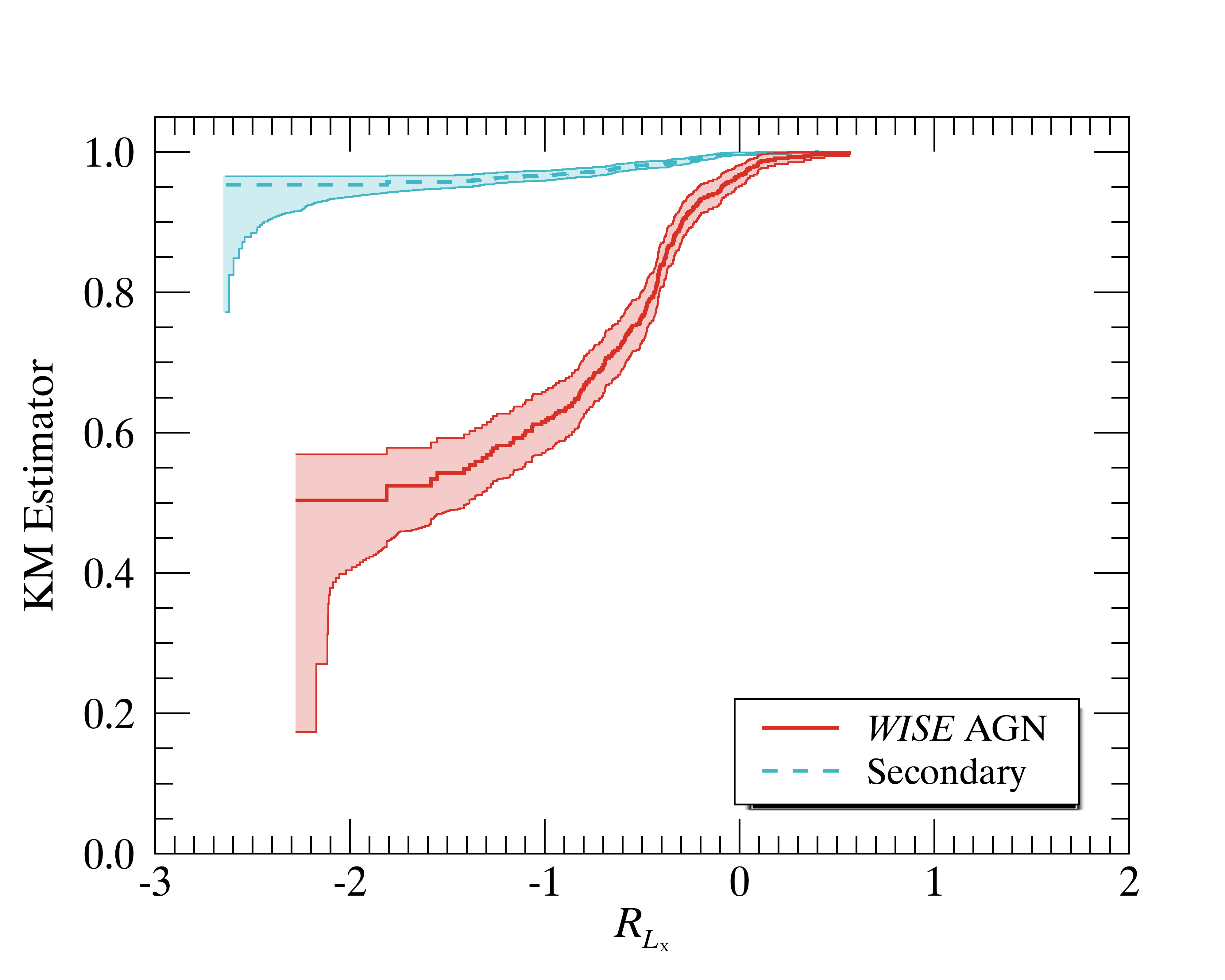}
\caption{Survival analysis using the KM estimator for both the \wise AGNs and secondary sources in our analysis subset. The KM estimator is interpreted as a cumulative distribution of sources with increasing \rlum. The lack of X-ray detections at low values of \rlum leads to increased uncertainty in the estimator at values lower than \mbox{\rlum = $-$2.0}.
\label{km_est}}
\end{figure}

However, a direct comparison of X-ray detected sources to upper limits could bias our results and lead to incorrect assumptions of the underlying population of heavily obscured AGNs. To account for our X-ray non-detected sources, we performed a survival analyses on our sample to understand the effects of the censored data on our \nh estimates. We used a Kaplan-Meier estimator (KM) to determine the survival function for both the \wise AGNs and our secondary sources, given increasing values of \rlum (see Figure \ref{km_est}). The \nqwac \wise AGNs in our analysis subset consists of \nqdetwac X-ray detections and \nqnonwac non-detections. Similarly, the \nqres secondary sources consist of \nqdetres X-ray detections and \nqnonres non-detections. While the number of X-ray non-detected to detected \wise AGNs are comparable (\mbox{{$\sim$}60\%}), the censoring rate of our secondary sources is extremely high (\mbox{{$\sim$}97\%}). Furthermore, at low X-ray luminosity ratios (\mbox{\rlum $\lesssim -$2.0}), all sources are censored for both the \wise AGNs and secondary sources. The disproportionate number of X-ray non-detections and lack of uncensored data at low \rlum affects the survival analysis and eliminates any reliable inference from the KM estimator at these values.

\begin{figure}
\epsscale{1.1}
\plotone{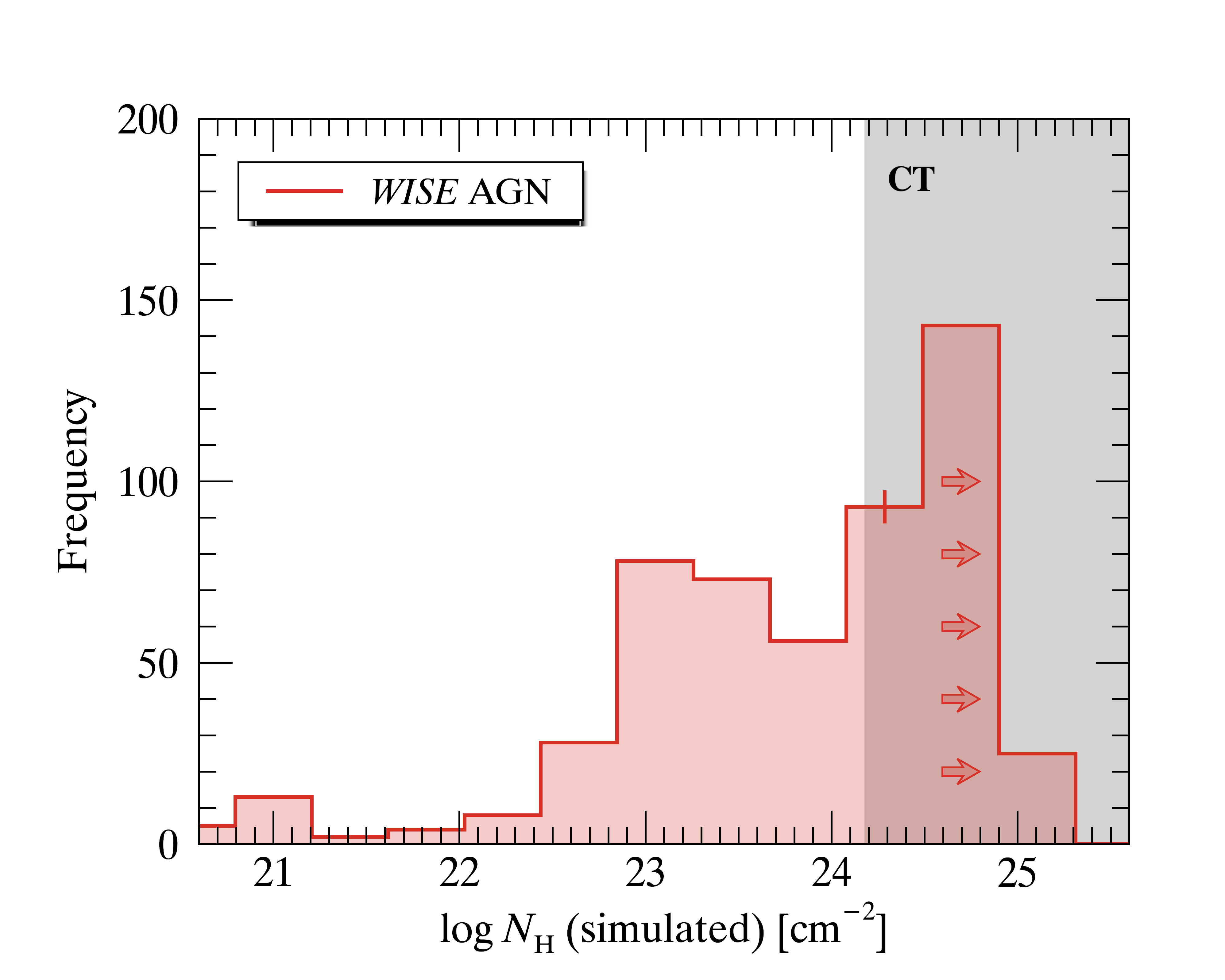}
\caption{Simulated \nh distribution drawn from the differential KM estimator for the \wise AGNs in our sample, using the \textsc{borus} model of nuclear X-ray emission. The vertical tick mark demarcates the \nh bin corresponding to the minimum \rlum of uncensored data in our survival analysis. As only censored observations exist beyond this point, these values are in turn lower limits.
\label{sim_nh}}
\end{figure}

In this form, the KM estimator is interpreted as the cumulative distribution of X-ray detected sources with increasing \rlum. We calculated the differential of the KM estimator (see \citealt{wardle1986} for details) and evenly distributed the remaining censored sources beyond where our uncensored data terminate (i.e., bins where \rlum $\lesssim -$2.0). We then uniformly distributed the sources within each bin of \rlum to reconstruct the \rlum distribution for our analysis subset---now simulated by the survival function---and converted the X-ray luminosity ratios to \nh using the \textsc{borus} model as described in Section \ref{nh_est} (see Figure \ref{sim_nh}.) Although our survival analysis is indeterminate at excessively low \rlum, it is significant to and beyond the X-ray luminosity ratio corresponding to Compton-thick levels of obscuration (\mbox{\rlum = $-$1.567}).

Additionally, we subtracted the known X-ray detected \rlum distribution from the simulated distribution, leaving us with a distribution in \rlum for strictly X-ray non-detected sources. We then sampled from this distribution to infer a simulated \nh distribution for the X-ray non-detected sources (see Figure \ref{sim_nh}).

This result provides additional support for the need of a large population of CT AGNs to explain the black hole mass density while adhering to constraints imposed by the cosmic X-ray background (\citealt{comastri2015}). It should be noted the distributions presented here are derived from modeled X-ray data with certain assumptions (e.g., source redshift, photon index, lack of intervening dust on larger scales), and that a more thorough treatment is required to understand the underlying \nh distributions. A detailed analysis of such modeling is left for future work.

\subsection{X-ray Stacking Analysis}
As an additional test of our ability to select heavily obscured AGNs with no X-ray counterparts, we performed an X-ray stacking analysis using \textsc{stackfast}\footnote{\url{http://www.dartmouth.edu/~stackfast/}}. For this analysis we chose to focus strictly on sources within the \chandra footprint, as \chandra yields the highest SNR of our three X-ray observatories. The \textsc{stackfast} software uses pre-reduced \chandra ACIS data products based on individual observations to match against a set of input sources. The pre-reduction stage allows \textsc{stackfast} to work much more efficiently than other stacking methods, as the data is only sorted and screened once, at initialization; the result of which is a set of stackable X-ray events containing position, energy, grade, and exposure times for each source within the master set of \chandra observations. For each one of the sources in our sample, we extracted photon counts and exposure times in both soft (\mbox{0.5--2 keV}) and hard (\mbox{2--7 keV}) energies, and estimated X-ray fluxes and luminosities (see Table \ref{stack_results}). Fluxes are computed using flux to count rate ratios characteristic of \chandra ACIS-I responses in Cycle 12 (in approximately the middle time frame of the observations in our stacking analyses), and assuming a power law spectrum with tyipcal Galactic absorption (\mbox{\nh = 10$^{\text{20}}$ cm$^{-{\text{2}}}$}) and photon index \mbox{$\Gamma$ = 1.4}. This spectral shape is broadly consistent with the observed ratios of hard and soft sources counts in our stacking analyses as shown in Table \ref{stack_results}. The counts to flux ratios adopted are \mbox{1.11$\times$10$^{-{\text{11}}}$ erg cm$^{-{\text{2}}}$ count$^{-{\text{1}}}$} (\mbox{0.5--2 keV} count rate to \mbox{0.5--2 keV} flux) and \mbox{3.04$\times$10$^{-{\text{11}}}$ erg cm$^{-{\text{2}}}$ count$^{-{\text{1}}}$} (\mbox{2--7 keV} count rate to \mbox{2--10 keV} flux). We obtained uncertainties in the stacked flux by bootstrap resampling of the input sources. This yields uncertainties on the average flux for the sample, which reflect the distribution in input fluxes and are typically larger than Poisson photon counting uncertainties. Additional details regarding \textsc{stackfast} can be found in earlier implementations of the code (e.g., \citealt{hickox2007b,chen2013,goulding2017}).

\begin{deluxetable*}{lrrrrrrr}
\tablecolumns{8}
\tablecaption{X-ray stacking data and results.\label{stack_results}}
\tablehead{
\colhead{} & \colhead{$t_{\text{exp}}$} & \colhead{Energy} & \colhead{$N_{\text{src}}$} & \colhead{$N_{\text{bg}}$} & \colhead{$N_{\text{net}}$} & \colhead{$F_{\text{X}}$} & \colhead{$\log L_{\text{X}}$} \\
\colhead{} & \colhead{[Ms]} & \colhead{[keV]} & \colhead{} & \colhead{} & \colhead{} & \colhead{[erg\,s$^{-1}$\,cm$^{-2}$]} & \colhead{[erg\,s$^{-1}$]}
}
\startdata
\multirow{2}{*}{\wise AGN (X-ray det.)}     & \multirow{2}{*}{ 1.34} & 0.5--2 & 7881 &   91 & 7789 & (3.61$\pm$1.00)$\times10^{-14}$ & 43.41$\pm$0.12 \\
                                            &                        & 2--7   & 3856 &  118 & 3737 & (8.45$\pm$1.99)$\times10^{-14}$ & 43.78$\pm$0.10 \\
                                                                     \cline{2-8}
\multirow{2}{*}{\wise AGN (X-ray non-det.)} & \multirow{2}{*}{ 1.48} & 0.5--2 &  118 &   22 &   95 & (3.99$\pm$1.18)$\times10^{-16}$ & 41.54$\pm$0.13 \\
                                            &                        & 2--7   &  117 &   52 &   64 & (1.32$\pm$0.36)$\times10^{-15}$ & 42.05$\pm$0.12 \\
                                                                     \cline{2-8}
\multirow{2}{*}{Secondary (X-ray det.)}     & \multirow{2}{*}{ 1.18} & 0.5--2 & 1516 &   71 & 1444 & (7.60$\pm$2.20)$\times10^{-15}$ & 42.84$\pm$0.13 \\
                                            &                        &   2--7 & 1256 &  110 & 1145 & (2.94$\pm$0.64)$\times10^{-14}$ & 43.42$\pm$0.09 \\
                                                                     \cline{2-8}
\multirow{2}{*}{Secondary (X-ray non-det.)} & \multirow{2}{*}{22.57} & 0.5--2 & 1190 &  801 &  388 & (1.07$\pm$0.14)$\times10^{-16}$ & 40.78$\pm$0.05 \\
                                            &                        &   2--7 & 1210 &  972 &  237 & (3.20$\pm$0.61)$\times10^{-16}$ & 41.25$\pm$0.08 \\
                                                                     \cline{2-8}
\multirow{2}{*}{Removed AGN}                & \multirow{2}{*}{18.51} & 0.5--2 & 1108 &  672 &  435 & (1.46$\pm$0.14)$\times10^{-16}$ & 40.84$\pm$0.04 \\
                                            &                        & 2--7   & 1076 &  847 &  228 & (3.76$\pm$0.80)$\times10^{-16}$ & 41.25$\pm$0.09 \\
                                                                     \cline{2-8}
\multirow{2}{*}{SED galaxy}                 & \multirow{2}{*}{21.67} & 0.5--2 & 1301 &  858 &  442 & (1.27$\pm$0.13)$\times10^{-16}$ & 40.78$\pm$0.05 \\
                                            &                        &   2--7 & 1023 &  970 &   52 & (7.39$\pm$5.63)$\times10^{-17}$ & 40.54$\pm$0.33 \\
\enddata
\tablecomments{X-ray stacking results for energy ranges of \mbox{0.5--2} and \mbox{2--7 keV}. For each energy range, we present exposure time, photon counts, and flux and luminosity estimates. The flux and luminosity values listed in the \mbox{2--7 keV} energy range have been scaled to \mbox{2--10 keV} for direct comparison to the body of this work.}
\end{deluxetable*}

\begin{figure}
\epsscale{1.1}
\plotone{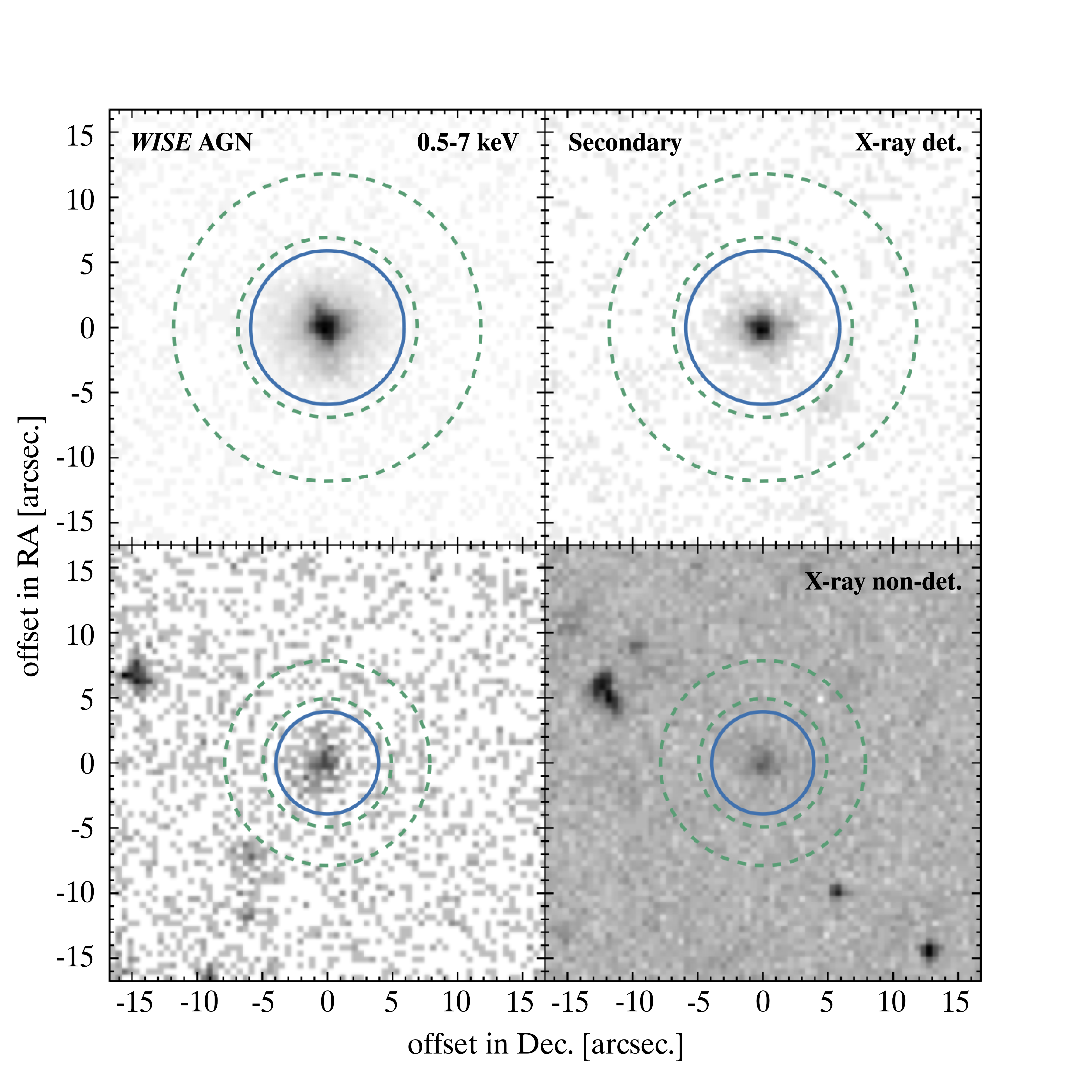}
\caption{X-ray stacking results within the \chandra footprint of the \wise AGN (\emph{left column}) and secondary sources (\emph{right column}), for both the X-ray detected (\emph{top row}) and X-ray non-detected (\emph{bottom row}) sources. In each panel, the solid, blue circle highlights the source extraction region, while the dashed, green circles depict the annulus of the background subtraction region. See Table \ref{stack_results} for corresponding data.
\label{xstack}}
\end{figure}

We produced stacked images for our \wise AGN and secondary sources, further grouped by X-ray detected and non-detected sources (see Figure \ref{xstack}). Evidence of X-ray emission above the background is clearly visible in all four stacked images, including the sources not individually detected by \chandra (bottom row of Figure \ref{xstack}). In the hard X-ray energy range, the X-ray detected stacks have luminosities on the order of \mbox{42 $\lesssim {\text{log \lx}} \lesssim$ 44}, while the X-ray non-detected stacks have luminosities on the order of \mbox{41 $\lesssim {\text{log \lx}} \lesssim$ 42}. For the \wise AGN stacks and the secondary X-ray detected stack, the estimated luminosities are within reasonable AGN luminosity ranges. The secondary X-ray non-detected stack also exhibits enough power to be considered a low-luminosity AGN (e.g., \citealt{fornasini2018}), more than expected for typical star-forming galaxies. Given the estimated X-ray luminosity in Table \ref{stack_results}, the secondary non-detected sample would require star-formation rates on the order of \mbox{{$\sim$}35 $M_\Sun$ yr$^{-{\text{1}}}$} (see Figure 16 of \citealt{lehmer2019}).

As a final analysis, we compared the X-ray stacking results of our X-ray non-detected secondary sources to both the candidate AGNs removed from our analysis subset in Section \ref{sec:subset}, as well as the sources removed for lack of AGN contribution to their SEDs (i.e., SED galaxies of Table \ref{qual_cut}). We compared the different sample groupings in both soft and hard X-rays. At softer energies, the estimated X-ray fluxes of the three groups are nearly identical. At harder energies, the difference between the AGN and galaxy groupings become apparent. For both the secondary and removed AGN samples, the hardness ratio remains comparable, but is much softer for the SED galaxy subset, providing further evidence that our SED modeling and selection criteria accurately identifies AGN activity.

\begin{figure}
\epsscale{1.1}
\plotone{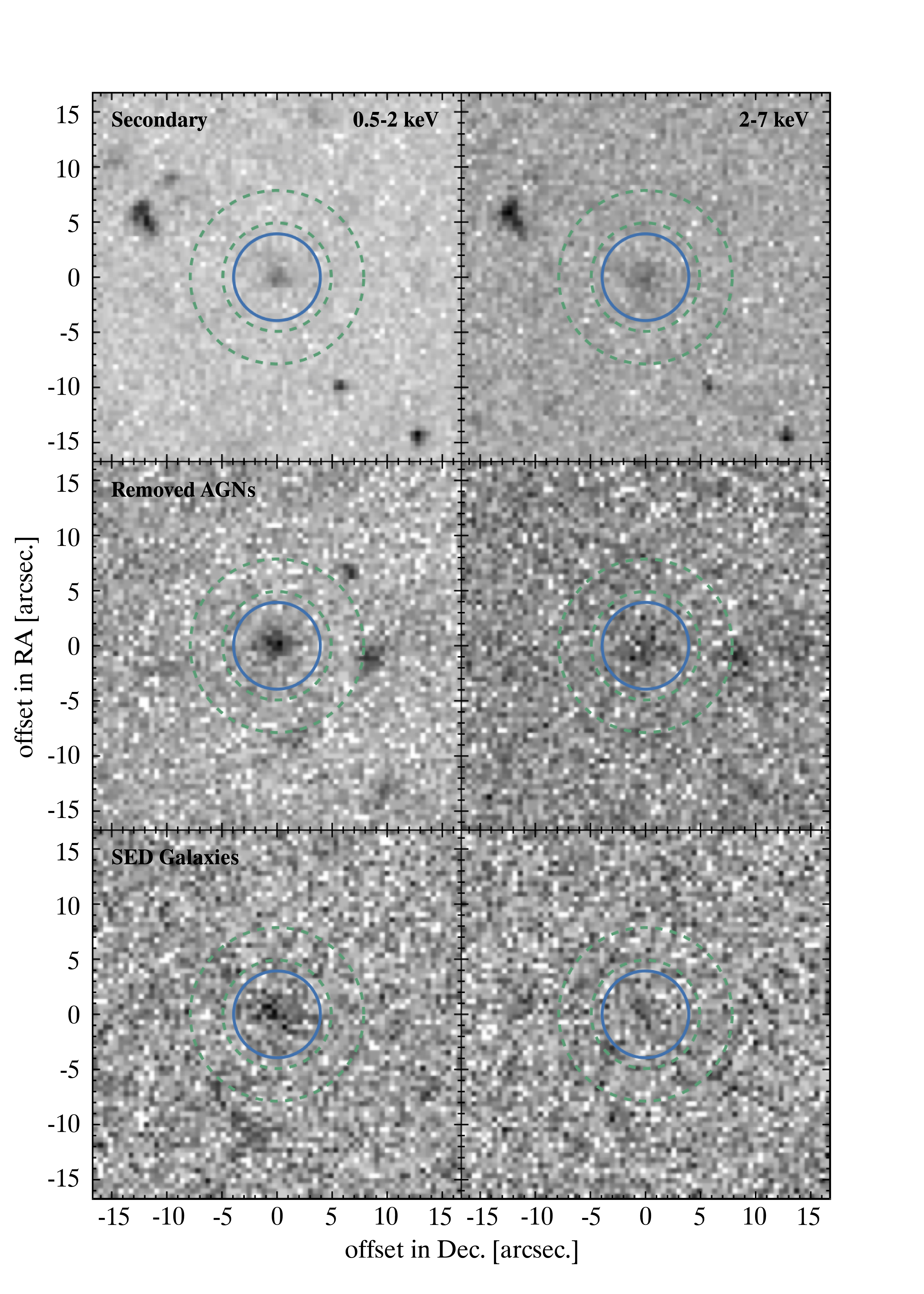}
\caption{X-ray stacking results for X-ray non-detections of our secondary sources, compared to the candidate AGNs removed from our analysis subset and sources with only galaxy contribution to their SEDS (see Section \ref{sec:subset}). In each panel, the solid, blue circle highlights the source extraction region, while the dashed, green circles depict the annulus of the background subtraction region. A clear signal in both the removed AGNs and SED galaxies can be seen in soft X-ray stacks (\mbox{0.5--2 keV}; \emph{left column}). In harder X-rays (\mbox{2--7 keV}; \emph{right column}), the signal remains for the removed AGNs, but is not present in the SED galaxies. See Table \ref{stack_results} for corresponding data.
\label{gstack}}
\end{figure}


\section{Discussion and Conclusions}
\label{sec:discussion}
In this paper, we presented a large sample of candidate AGNs with strong MIR emission, lacking X-ray counterparts in observed \chandra, \xmm, or \nustar observations. We calculated observed-to-intrinsic X-ray luminosity ratios \rlum, with intrinsic X-ray luminosities inferred from the \lx--\lir relation given AGN MIR luminosities from SED modeling. We substituted X-ray flux limits in place of detections for sources without X-ray counterparts and calculated upper limits on \rlum for said sources. The results of our SED modeling and X-ray luminosity ratio calculations show a strong anticorrelation between \rlum and \ebv. Using multiple X-ray spectral models, we estimated obscuring column densities, converting \rlum to \nh. The results of a survival analysis between our X-ray detected and non-detected sources provides a clear indication for a population of highly obscured sources reaching CT levels of obscuration, and is consistent with the results of our X-ray stacking analysis. To more accurately understand the underlying distribution of extremely obscured AGNs, future work must be done to forward model the \nh distribution to understand our observable variables, namely \rlum.

Strong IR emission from AGNs is a clear indication for the existence of dense obscuring material within the nuclear region (e.g., \citealt{elitzur2006}), as well as total AGN power. While X-ray emission is mostly resistant to obscuration at moderate levels (\nh $<$ 10$^{\text{22}}$ \cmcm), very large column densities, up to and above the CT regime, can conceal even the hardest X-ray energies (e.g., \citealt{yan2019}), potentially making these AGNs undetectable with typical X-ray observations on current facilities. The relation between X-ray and IR emission in AGNs provides an approach for estimating the intrinsic X-ray luminosity of a source, regardless of obscuration. Thus, the ratio of observed-to-intrinsic X-ray luminosity (\rlum) can be used as a proxy for nuclear obscuration.

Understanding the full AGN population, both obscured and unobscured, is crucial to answering many research questions such as the observed fraction of obscured AGNs, their contribution to the Cosmic X-ray Background (CXB), and constraining the luminous end of the AGN luminosity function that is not directly probed in X-ray surveys. Furthermore, the AGN contribution to galaxy emission (as shown in our SED models) is a proxy for AGN accretion and thus BH growth (i.e., \citealt{soltan1982}). Constraints imposed on the AGN luminosity function from heavily obscured AGNs can be compared to the local BH mass function (\citealt{lauer2007}). Such work could map the accretion history of the universe and determine the radiative efficiency of accretion (e.g., \citealt{merloni2008}) with the addition of a population of heavily obscured AGNs that have not yet been identified in X-rays.


\acknowledgments
We thank the referee for invaluable comments which lead to a significant improvement in this work.

This research has made use of X-ray data from the following: the \nustar mission, a project led by the California Institute of Technology, managed by the Jet Propulsion Laboratory, and funded by the National Aeronautics and Space Administration (NASA) (NASA); the \chandra Source Catalog, provided by the \chandra X-ray Center (CXC) as part of the \chandra Data Archive; XMM-Newton, a European Space Agency (ESA) science mission with instruments and contributions directly funded by ESA Member States and NASA. This research also made use of data products from the \emph{Wide-field Infrared Survey Explorer}, which is a joint project of the University of California, Los Angeles, and the Jet Propulsion Laboratory/California Institute of Technology, and funded by NASA. This work is also based in part on NIR data obtained as part of the UKIRT Infrared Deep Sky Survey, as well as data products from the Two Micron All Sky Survey, which is a joint project of the University of Massachusetts and the Infrared Processing and Analysis Center/California Institute of Technology, funded by NASA and the National Science Foundation.

This research also made use of software provided by the High Energy Astrophysics Science Archive Research Center (HEASARC), which is a service of the Astrophysics Science Division at NASA/GSFC and the High Energy Astrophysics Division of the Smithsonian Astrophysical Observatory.

C.M.C. and R.C.H. acknowledge support from the NSF through grant number 1554584. C.M.C. acknowledges support from the NASA New Hampshire Space Grant Consortium, and from a Dartmouth Fellowship. R.C.H. acknowledges support from NASA through grant numbers NNX15AP24G, 18-ADAP18-0105, and Chandra GO award GO7-18130X.

R.J.A. was supported by FONDECYT grant number 1191124.

\facilities{\chandra, \galex, MMT, \nustar, SDSS, UKIRT, \wise, \xmm, 2MASS}

\clearpage
\bibliography{ccarroll}

\end{document}